\documentclass[journal,twocolumn]{IEEEtran}

\usepackage{multicol}
\usepackage{multirow}
\usepackage{graphicx}
\usepackage[fleqn]{amsmath}
\usepackage{algorithm}
\usepackage[labelsep=space]{caption}
\usepackage{setspace}
\usepackage{algpseudocode}
\usepackage{amssymb}
\usepackage{bm}
\usepackage{color}
\usepackage{colortbl}
\usepackage{amsmath}
\usepackage{cases}
\usepackage{amssymb}
\usepackage{array}
\usepackage{booktabs}
\usepackage{hyperref}
\usepackage[tight,footnotesize]{subfigure}
\usepackage{enumerate}
\usepackage{mathtools,cuted}
\usepackage{stfloats}

\usepackage{cite}
\usepackage{subfigure}
\usepackage{float}
\usepackage{multirow}
\usepackage{caption}
\usepackage[justification=centering]{caption}
\usepackage{caption}
\usepackage{url}
\usepackage{leftidx}
\usepackage{makecell}
\ifCLASSINFOpdf
\else
\fi

\begin{document}
\title{Track-before-detect Algorithm based on Cost-reference Particle Filter Bank for Weak Target Detection}

\author{Jin~Lu, Guojie Peng, Weichuan~Zhang,~\IEEEmembership{Member,~IEEE}, Changming Sun
\thanks{This work was supported by the National Natural Science Foundation of China under Grant 61801281.}% <-this % stops a space
\thanks{Jin Lu, Guojie Peng and Weichuan Zhang are with School of Electrical Information and Artificial Intelligence, Shaanxi University of Science \& Technology, Xi'an, 710021, China e-mail: lujin@sust.edu.cn, lj491216@163.com, zwc2003@163.com.}% <-this % stops a space

\thanks{Changming Sun is with CSIRO Data61, PO Box 76, Epping, NSW 1710, Australia email: changming.sun@csiro.au.}
}

%\markboth{IEEE Transactions on Signal Processing,~Vol.~14, No.~8, November~2020}%


\maketitle

\begin{abstract}
Detecting weak target is an important and challenging problem in many applications such as radar, sonar etc.
However, conventional detection methods are often ineffective in this case because of low signal-to-noise ratio (SNR).
This paper presents a track-before-detect (TBD) algorithm based on an improved particle filter, i.e. cost-reference particle filter bank (CRPFB), which turns the problem of target detection to the problem of two-layer hypothesis testing.
The first layer is implemented by CRPFB for state estimation of possible target.
CRPFB has entirely parallel structure, consisting amounts of cost-reference particle filters with different hypothesized prior information.
The second layer is to compare a test metric with a given threshold, which is constructed from the output of the first layer and fits GEV distribution.
The performance of our proposed TBD algorithm and the existed TBD algorithms are compared according to the experiments on nonlinear frequency modulated (NLFM) signal detection and tracking.
Simulation results show that the proposed TBD algorithm has better performance than the state-of-the-arts in detection, tracking, and time efficiency.
\end{abstract}


% \fi
\IEEEpeerreviewmaketitle
\section{Introduction}
\IEEEPARstart{D}{etecting} weak target with low signal to noise ration (SNR) is crucial in many applications.
However, conventional detect-before-track (DBT) approach~\cite{r1}-\cite{r2} is invalid in this case.
In DBT, firstly raw measurements are compared with a given threshold to detect signals and reduce data flow, then the following tracking step operates on these exacted measurements.
As a result, high threshold may lead to loss of potential information and low threshold may lead to high rate of false-alarm.

By contrast, track-before-detect (TBD) approach~\cite{davey2012using, grossi2013novel, feichtenhofer2017detect, yi2020multi, grossi2014track} is valid for low SNR target detection and tracking. In TBD approach, unthresholded or raw measurements are processed for jointly detection and tracking and thus the potential information of the possible targets is preserved.

TBD algorithms can be roughly divided into two categories: recursive algorithms and batch algorithms.
In recursive algorithms, the detection or existence probability is reported at each time step. Then the state estimate of the time step is given if a target is declared.
Particle filter (PF) is one kind of popular common algorithm to implement recursive TBD.
PF is a class of suboptimal Beyesian filter, in which the posterior probability distribution function is approximated by a large number of weighted particles or samples~\cite{djuric2003particle, candy2016bayesian, li2015resampling, nummiaro2003adaptive, gustafsson2010particle, van2000unscented, arulampalam2002tutorial, gustafsson2002particle}. Therefore, PF is proper to process nonlinear/non-Gaussian dynamic system.
In PF based TBD (PF-TBD), a discrete variable is added to state vector to mimic the presence or absence of a target. Detection and estimation results are given at each time instant. PF was first applied to TBD by Salmond in \cite{b22} and Boers in \cite{b23}. Then it was extended to multitarget tracking by Ristic and Rutten~\cite{ristic2003beyond,rutten2005recursive}.
We also proposed a recursive TBD algorithm based on cost-reference particle filter (CRPF)~\cite{lu2017cost}. CRPF is a new kind of PF, requiring none of statistical information of dynamic system ~\cite{b12}.

%In this paper, we focus on the second category TBD algorithms, i.e., the batch algorithms.
Compared with recursive algorithms, batch TBD algorithms accumulate measurements along the possible trajectories of the possible targets in an observed interval.
The cumulated energy is taken as a test metric to compare with a threshold. A target is declared if the test metric exceeds the given threshold.
Meanwhile, the corresponding trajectory is output if a target is declared.
Because of cumulating energy in the whole observed interval, batch TBD algorithms always provide more accuracy estimation and detection results.
Batch TBD algorithms are generally implemented using Hough transform~\cite{carlson1994search,carlson1994search2, zhao2021deep},
dynamic programming~\cite{johnston2002performance, zhu2022optimization}, maximum likelihood~\cite{tonissen1998maximum, yi2020multi}, etc.
However,these methods prohibit or penalize deviations from the straight-line motion, and in general require enormous computational resources~\cite{r2}.
%Yi et al.~\cite{yi2020multi} proposed a multi-frame TBD (MF-TBD) algorithm with measurement-directed strategy, which considers both the a prior information and measurement direction in state estimation.
We proposed a batch TBD algorithm based on cost-reference particle filter (CRPF-detector), which can detect and track low SNR target with unknown statistical information~\cite{lu2021cost}.
Shui et al.~\cite{shui2016detection} proposed a batch TBD algorithm based on forward-backward cost-reference particle filter (FB-CRPF-detector) for nonlinear frequency modulated (NLFM) signal detection in radar.
%Forward-backward CRPF is an improved CRPF~\cite{b12}, thus the method can deal with more complicated problem with time-varying velocity.
%In forward-backward CRPF based TBD algorithm (FBCRPF-TBD), a two-dimensional feature calculated from the output of backward CRPF is taken as a test metric, in which backward CRPF is initialized by the output of forward CRPF.

Although CRPF-detector and FB-CRPF-detector could detect and track the target with nonlinear motion, they still suffer from two problems.
The first is high computational load.
In FBCRPF-TBD, forward CRPF and backward CRPF are sequentially carried out and the number of particles required by both the CRPFs is large. Furthermore, the test metric of FBCRPF-TBD is two dimensional and implemented in an convex area.
The second is the unknown probability distribution of the test metric. These methods generally require numerous noisy-only measurements to estimate the detection threshold with given false alarm probability. The test metric of FBCRPF-TBD are two-dimensional. The statistical properties of both the methods are unknown, thus numerous noisy only measurements are required to estimate the detection threshold for given false alarm rate.

In this paper, we propose a batch TBD algorithm based on CRPF bank (CRPFB).
In proposed TBD algorithm, firstly CRPFB is designed to estimate the state sequence of possible target. As an improved particle filter, CRPFB consists of many paralleled CRPFs using different but precise, hypothesized prior information and the number of particles required by each CRPF is very small.
Therefore, the estimation accuracy and computational complexity of CRPFB are both improved.
Secondly a total test metric is used to determine the presence or absence of the target during the observation time.
The total test metric of proposed algorithm fits the generalized extreme value (GEV) distribution, thus the detection performance of the proposed method can be analytically estimated by using extreme value theory.

The main contributions of our proposed method comprise of three aspects.
Firstly, the impact of precise initial information on the performance of particle filtering methods is illustrated.
Secondly, an improved PF, i.e., CRPFB, with entirely paralleled structure is proposed.
We try to reduce the computational burden of particle filtering methods by transforming the problem of filtering to the problem of hypothesis testing.
Thirdly, a batch TBD algorithm based on CRPFB is proposed for low SNR target detection and tracking, of which the total test metric fits GEV distribution~\cite{b21, b32}. We try to explore an effective way to analytically analyze the detection performance of batch TBD algorithms.

The remainder of the paper is organized as follows.
In Section II, the research background related to the proposed method is introduced, including the dynamic system state of the detection and tracking problem, and the brief introduction of CRPF.
In Section III, firstly the impact of initialization on PF is revealed, and then the CRPFB is proposed.
In Section IV, the ways of detecting target over the whole observation interval and at each time step are shown, respectively. Moreover, the distribution of the overall test metric is analyzed.
In Section V, several simulations on moving target detection and tracking in low SNR are shown to illustrate the good performance of our proposed method.
A conclusion is given in Section VI.

\section{Background}
In this section, we show the background of the proposed algorithm, including the measurement model for multi-frame target detection, the piecewise constant velocity state space model for CRPFB, a brief introduction of original CRPF.

\subsection{measurement model for target detection}
The problem of target detection can be summarized as a binary hypothesis testing \cite{shui2016detection} as follows:
\begin{equation}\label{eq1}
  \begin{aligned}
  H_{1}: \textbf{z}_{t_{k}}&=f_{y}(\textbf{x}_{t_{k}})+\textbf{w}_{t_{k}}, k=1,2,\ldots,K \\
  H_{0}: \textbf{z}_{t_{k}}&=\textbf{w}_{t_{k}},\\
  \end{aligned}
\end{equation}
where $H_{1}$ denotes the hypothesis of target existence at time $t_{k}$ and $H_{0}$ denotes the hypothesis of target absence at time $t_{k}$. Under the hypothesis $H_{1}$, the observation vector $\textbf{z}_{t_{k}}$ at time $t_{k}$ depends on the system state $\textbf{x}_{t_{k}}$ and the observation noise $\textbf{w}_{t_{k}}$; under the hypothesis $H_{0}$, $\textbf{z}_{t_{k}}$ contains noise only.
$f_{y}$ is a transformation of the system state $\textbf{x}_{t_{k}}$, which can be linear or nonlinear.
The goal of target detection is to test the hypotheses $H_{0}$ and $H_{1}$.

\subsection{State-space model for target estimation by using CRPFB}
To solve the target detection problem by using TBD algorithm based on CRPFB,
a state-space model under the hypothesis $H_{1}$ is necessary.
In this paper, we construct a piecewise constant velocity state-space model for CRPFB, the state equation for which is shown in Equation~\ref{eqt2a}.
For convenience, target state vector $\textbf{x}_{t_{k}}$ is set as $\textbf{x}_{t_{k}}=[x_{t_{k}},\dot{x}_{t_{k}},y_{t_{k}},\dot{y}_{t_{k}}]^{\texttt{T}}$,
where $(x_{t_{k}},y_{t_{k}})$ denotes the location of the possible target at time $t_{k}$, $(\dot{x}_{t_{k}},\dot{y}_{t_{k}})$ is the velocity of the possible target at time $t_{k}$. For CRPFB, amplitude of echo signal is not necessary.

\begin{equation}\label{eqt2a}
  \begin{aligned}
  \begin{bmatrix}  x_{t_{k}}\\
                  \dot{x}_{t_{k}} \\
                  y_{t_{k}} \\
                  \dot{y}_{t_{k}}
  \end{bmatrix} = \begin{bmatrix} 1 & \triangle T & 0 & 0  \\
  0 & 1 & 0 & 0  \\
  0 & 0 &1 & \triangle T  \\
  0 & 0 & 0 & 1
  \end{bmatrix}
  \begin{bmatrix} x_{t_{k-1}} \\
  \dot{x}_{t_{k-1}} \\
  y_{t_{k-1}} \\
  \dot{y}_{t_{k-1}}
  \end{bmatrix}+ \begin{bmatrix}
                                     \frac{\triangle T}{2} & 0 \\
                                     1 & 0\\
                                     0 & \frac{\triangle T}{2} \\
                                     0 & 1
                                   \end{bmatrix} \begin{bmatrix}
                                                   v_{t_{k},1} \\
                                                   v_{t_{k},2}
                                                 \end{bmatrix},
  \end{aligned}
\end{equation}

It can be briefly shown as follows, where $\textbf{v}_{t}$ is the system noise with zero mean and covariance \textbf{v} and is independent from target states.
\begin{equation}\label{eqt2b}
  \begin{aligned}
  \textbf{x}_{t_{k}}=\textbf{A}\textbf{x}_{t_{k-1}}+\textbf{B}\textbf{v}_{t_{k}}.
  \end{aligned}
\end{equation}

With the measurement model in Equation~(\ref{eq1}) under hypothesis $H_{1}$, the state-space model for target estimation by using CRPFB is constructed.

\subsection{Cost-reference particle filter}
Based on the piecewise constant velocity state-space model in section II.B, a brief introduction of CRPF is present here, as the basis for CRPFB.

In \cite{b12}, M\'{\i}guez et al. proposed a new type of particle filtering method (i.e. CRPF, or generalized particle filter, GPF) that does not assume explicit mathematical forms of the probability distributions of the noise in the system. CRPF has the similar structure to PF. In PF, the distributions of the system noise and measurement noise are required to calculate particle weights and approximate the posterior probability density distribution of the state~\cite{b27}. Different from PF, CRPF does not need the statistical information of the state-space model, in which the particles are measured by the user-defined cost and the particle with the minimum cost is considered as the state estimation. The basic principle of state estimation of CRPF is shown as follows.

\begin{equation}\label{eq1a}
  \begin{aligned}
  &\textrm{particle-cost set}: \{\textbf{x}_{t_{k}}^{i},C_{t_{k}}^{i}\}_{i=1}^{N}, \\
  &\textrm{state estimation}: \hat{\textbf{x}}_{t_{k}}=\textbf{x}_{t_{k}}^{i_{\textrm{min}}},
  i_{\textrm{min}}=\min_{i}\{C_{t_{k},m}^{i}\}_{i=1}^{N},\\
  &or, \hat{\textbf{x}}_{t_{k}}=\sum_{i=1}^{N}\mu_{k}^{i}\textbf{x}_{t_{k}}^{i}.
  \end{aligned}
\end{equation}
where $\textbf{x}_{t_{k}}^{i}$ is the ${i}$-th particle at time $t_{k}$ with cost $C_{t_{k}}^{i}$ . The calculation of $C_{t_{k}}^{i}$ does not depend on statistical information of the system. The state estimation at time $t_{k}$ is the particle with smallest cost, or the weighted sum of all the particles, where $\mu_{t_{k}}^{i}$ is the \emph{weight} of particle $\textbf{x}_{t_{k}}^{i}$, and is generally a inverse proportional function of particle cost $C_{t_{k}}^{i}$.

%\subsection{Generalized extreme value distribution}
%The generalized extreme value (GEV) distribution is an important model of extreme value theory, which is the basic of the threshold estimation in the proposed TBD algorithm.
%
%The generalized extreme value (GEV) distribution is often used to model the smallest or the largest value among a large set of independent, identically distributed random values representing measurements or observations. Provided that there are batches of 1,000 points from a process, if the maxima of each batch is recorded, the data are known as block maxima (or minima if the minima is recorded). The generalized extreme value distribution can be used as a model for those block maxima.
%The generalized extreme value distribution is a generalization of the extreme value distribution, which combines three distributions into a single form, allowing a continuous range of possible shapes that includes all three of the simpler distributions. The three distributions belong to the Gumbel, Fr\'{e}chet, and Weibull framework. The generalized extreme value distribution allows one to `let the data decide' which distribution is appropriate. It is worth to note that the three types of extreme value distributions are the only possible limitations for the distributions of block maxima or block minima, regardless of the distribution of the data \cite{30}, \cite{33}.

\section{Proposed Algorithm}
In this section, we present the batch TBD algorithm based on CRPFB (CRPFB-TBD) for low SNR target detection and tracking.
The proposed algorithm includes two steps:
first, an improved cost-reference particle filter for estimating the state sequence of a possible target;
second, a total test metric based on the estimate results of CRPFB for batch detection.

\subsection{Feature of piecewise constant velocity state-space model}
The design of CRPFB is based on the feature of piecewise constant velocity state-space model.
In this part, we show this feature and compare the hypothesised prior information based on this feature and the original initial information.

It is readily obtained from the system model in Equation (\ref{eqt2a}) that the mean $\bar{\dot{x}}$ of the velocity $\dot{x}_{t_{k}}$ is constant as illustrated in Equation~(\ref{eq4}).
\begin{equation}\label{eq4}
\begin{aligned}
  &\dot{x}_{t_{k}}=\dot{x}_{t_{k-1}}+v_{t_{k-1},1},\\
  &E(\dot{x}_{t_{k}})=E(\dot{x}_{t_{k-1}})=\bar{\dot{x}}.
\end{aligned}
\end{equation}

Moreover, based on the approximation of the piecewise constant velocity signal, $\bar{\dot{x}}$ can be estimated as follows.
\begin{equation}\label{eq5}
\begin{aligned}
  \bar{\dot{x}}&=\frac{1}{K}\sum_{k=1}^{K}E(\dot{x}_{t_{k-1}})\\
  &=\frac{1}{K}\sum_{k=1}^{K}\frac{E(x_{t_{k}}-x_{t_{k-1}})}{\triangle T}\\
  &=\frac{E(x_{t_{1}}-x_{t_{0}}+\ldots+x_{t_{K}}-x_{t_{K-1}})}{T}\\
  &=\frac{E(x_{t_{K}}-x_{t_{0}})}{T}.
\end{aligned}
\end{equation}

Equation (\ref{eq5}) indicates that the mean $\bar{\dot{x}}$ of velocity can be estimated from the location at the initial and last time steps by using the approximation of the piecewise constant velocity signal.

%%初始信息突然出现，应说明为什么采用这样的初始信息。与观测区域相关。
Because of the field of view (FOV) of observation units, the possible range of initial information of $x_{t_{k}}$ can be $x_{t_{k}}\in[x_{\text{min}},x_{\text{max}}]$, where $x_{\text{min}}$, $x_{\text{max}}$ denotes the minima and maxima of FOV in x-direction.
Therefore, $x_{t_{0}},x_{t_{K}}\in[x_{\text{min}},x_{\text{max}}]$, and the possible range of $\bar{\dot{x}}$ can be estimated as follows.

\begin{equation}\label{eq6}
 \frac{x_{\textrm{min}}-x_{\textrm{max}}}{T}\leq\bar{\dot{x}}\leq\frac{x_{\textrm{max}}-x_{\textrm{min}}}{T}.
\end{equation}

Moreover, if we set $x_{t_{0}}=x_{\text{val}}$ as an exact value, the possible range of $\bar{\dot{x}}$ can be re-estimated as follows.
\begin{equation}\label{eq7a}
 \frac{x_{\textrm{min}}-x_{\text{val}}}{T}\leq\bar{\dot{x}}\leq\frac{x_{\textrm{max}}-x_{\text{val}}}{T}.
\end{equation}
where $x_{\text{val}}\in [x_{\textrm{min}},x_{\text{max}}]$.

It can be seen from Equation (\ref{eq6}) that the possible range of $\bar{\dot{x}}$ is $2\frac{|x_{\text{max}}-x_{\text{min}}|}{T}$.
By contrast, the possible range of $\bar{\dot{x}}$ in Equation (\ref{eq7a}) is only $\frac{|x_{\text{max}}-x_{\text{min}}|}{T}$ and is half of that in Equation (\ref{eq6}).
\begin{figure*}[!t]
\setlength{\abovecaptionskip}{0pt}
\setlength{\belowcaptionskip}{0pt}
\centering
\includegraphics[width=7in,angle=0]{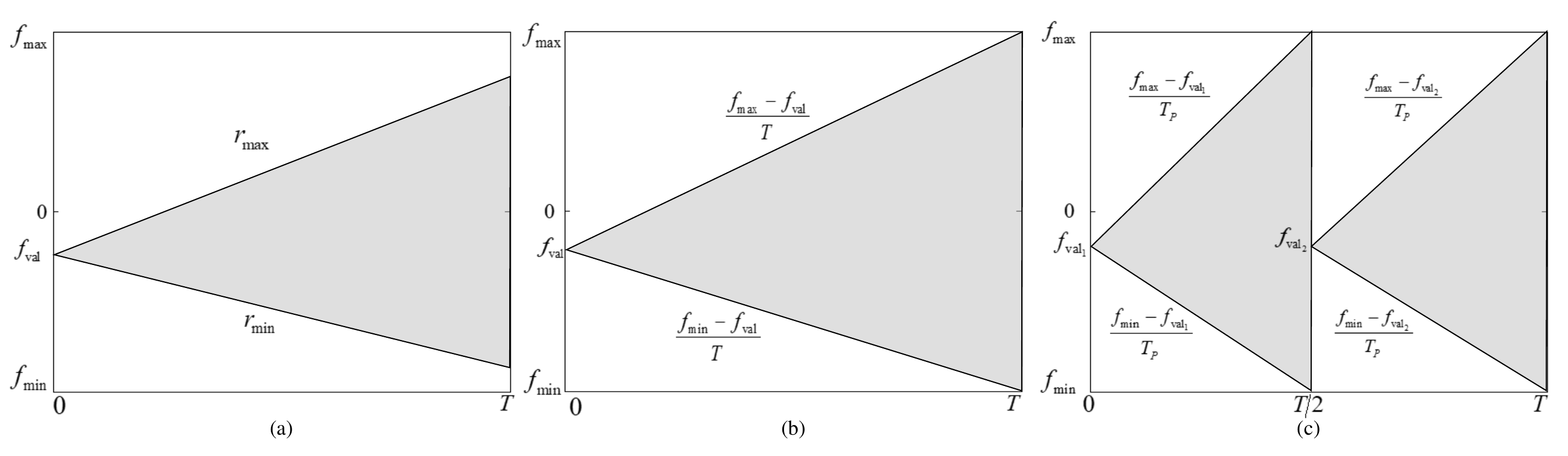}
\caption{Comparison of original prior information and hypothetical prior information, (a) Condition 1, (b) Condition 2, (c) Condition 3, with $T_{P}=0.5T$.}
\label{fig1}
\end{figure*}

It is reasonable to approximate the possible range of $\bar{\dot{x}}$ as the possible range of $\dot{x}_{t_{k}}$.
If the possible range of $\dot{x}_{t_{k}}$ is given as $\dot{x}_{t_{k}}\in[f_{\text{min}},f_{\text{max}}]$, there are three conditions as follows.
\begin{enumerate}[1]
  \item $\frac{|x_{\text{max}}-x_{\text{min}}|}{T}\gg \textrm{max}(|\dot{x}_{\text{min}}|,|\dot{x}_{\text{max}}|)$. \\
In this case, target returns are close to constant velocity (CV) signal. Supposed that $x_{t_{0}}=x_{\text{val}}$, and the corresponding possible range of $x_{t_{k}}$ can be as follows. However, we will not pay more attention on this simple case as CV problem in this paper.
\begin{equation}\label{eq7}
  \begin{aligned}
  &x_{t_{0}}=x_{\text{val}},\\
  &\dot{x}_{\textrm{min}}\leq {\dot{x}}_{t_{k}} \leq \dot{x}_{\textrm{max}},\\
  &x_{\text{val}}+\Delta T(k-1)\dot{x}_{\textrm{min}}\leq x_{t_{k}}\leq
  x_{\text{val}}+\Delta T(k-1)\dot{x}_{\textrm{max}}.
  \end{aligned}
\end{equation}
  \item $\frac{|x_{\textrm{max}}-x_{\textrm{min}}|}{T}\approx \textrm{max}(|\dot{x}_{\textrm{min}}|,|\dot{x}_{\textrm{max}}|)$.\\
In this case, target returns can be well modelled as a piecewise constant velocity signal. Supposed that $x_{t_{0}}=x_{\text{val}}$, and the corresponding possible range of $x_{t_{k}}$ can be as follows.
\begin{equation}\label{eq8}
  \begin{aligned}
  &x_{t_{0}}=x_{\text{val}},\\
  &\frac{x_{\textrm{min}}-x_{\text{val}}}{T}\leq {\dot{x}}_{t_{k}} \leq \frac{x_{\textrm{max}}-x_{\text{val}}}{T},\\
  &x_{\text{val}}+\Delta T(k-1)\frac{x_{\textrm{min}}-x_{\text{val}}}{T}\leq x_{t_{k}}
  \leq \\
  &x_{\text{val}}+\Delta T(k-1)\frac{x_{\textrm{max}}-x_{\text{val}}}{T}.
  \end{aligned}
\end{equation}
  \item $\frac{|x_{\textrm{max}}-x_{\textrm{min}}|}{T}\ll \textrm{max}(|\dot{x}_{\textrm{min}}|,|\dot{x}_{\textrm{max}}|)$.\\
In this case, during the observation time, the possible signal is severely nonlinear and cannot be suitably approximated as a piecewise constant velocity signal directly. One of the solutions is to shorten the observation time. We divide the observation time into equal length blocks $[0,T_{p}],[T_{p},2T_{p}],\ldots,[(P-1)T_{p},PT_{p}]$, $PT_{p}=T$, $\frac{x_{\textrm{max}}-x_{\textrm{min}}}{T_{p}}\approx \textrm{max}\{|\dot{x}_{\textrm{min}}|,|\dot{x}_{\textrm{max}}|\}$.
Filtering is implemented in each block.
For the $p$-th block, supposed that $x_{p,0}=x_{\text{val}}$, where $p=1,\cdots,P$, $f_{p,0}$ is the initial frequency of the $p$-th block, $x_{\text{val}}\in [x_{\text{min}},x_{\text{max}}]$.
\begin{equation}\label{eq9}
  \begin{aligned}
  &x_{p,0}=x_{\text{val}},\\
  &\frac{x_{\textrm{min}}-x_{\text{val}}}{T_{p}}\leq {\dot{x}}_{p,k} \leq \frac{x_{\textrm{max}}-x_{\text{val}}}{T_{p}},\\
  &x_{\text{val}}+\Delta T(k-1)\frac{x_{\textrm{min}}-x_{\text{val}}}{T_{p}}\leq x_{p,k}
  \leq \\
  &x_{\text{val}}+\Delta T(k-1)\frac{x_{\textrm{max}}-x_{\text{val}}}{T_{p}}.
  \end{aligned}
\end{equation}
\end{enumerate}
where $x_{p,k}$ is the left location in x-direction of the $k$-th subsequence of the $p$-th block.

The similar results can be obtained in y-direction.

Fig.~1 shows the original prior information $x_{t_{k}}\in[x_{\textrm{min}},x_{\textrm{max}}]$ (rectangles) and the prior information in Equations (\ref{eq7}), (\ref{eq8}), or (\ref{eq9}) determined by $x_{t_{0}}=x_{\text{val}}$ or $x_{p,0}=x_{\text{val}}$ and piecewise constant velocity signal model (shaded parts). It can be observed from Fig.~1 that the possible range of target location in Equations (\ref{eq7}), (\ref{eq8}), and (\ref{eq9}) is much smaller than the original prior information.
For PFs, the more precise prior information represented by the shaded parts indicates fewer number of particles, and much less computational complexity.

\subsection{Structure of CRPFB}
%% 修改图2中的初始信息符号，f-x,r-x'
Based on analysis in section III. A,  CRPFB is presented and its structure is shown in Fig.~2.
In Fig.~2, the process in the dashed box is CRPFB, $x_{t_{0},m}\sim U[x_\text{min},x_\text{max}]$ ($U[a, b]$ means uniform distribution among range $[a,b]$) is the exact value assigned to $x_{t_{0}}$ of the $m$-th CRPF, CRPF-$m$ dentoes the $m$-th CRPF, CRPF denotes the common cost-reference particle filter. $\hat{\textbf{\emph{X}}}_{m}$ is the estimated results provided by the $m$-th CRPF, $C_{\text{cum}}^{m}$ is the cumulated cost of the $m$-th CRPF, where $m=1,2,\cdots, M$, $m_{min}$ is the index of the minimum cumulated cost, $\hat{\textbf{\emph{X}}}_{m_\text{min}}$ is filtering results of CRPFB, $C_{\text{cum}}^{\text{min}}$ is the minimum cumulated cost.
\begin{figure*}[!b]
\setlength{\abovecaptionskip}{0pt}
\setlength{\belowcaptionskip}{0pt}
\centering
\includegraphics[width=6in,angle=0]{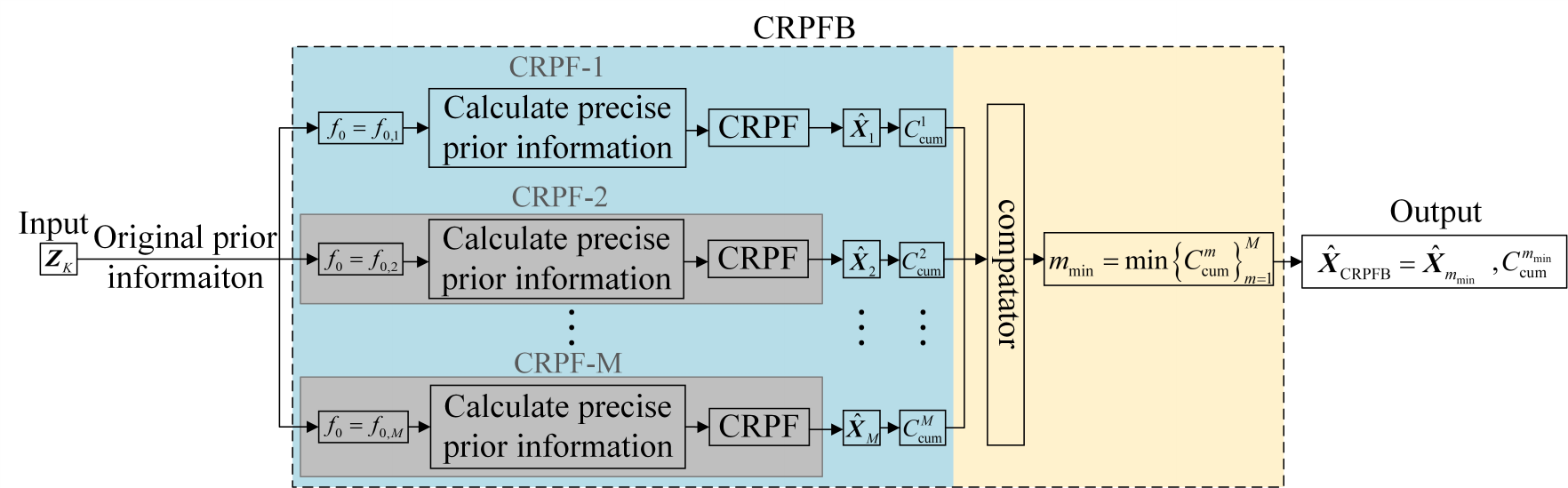}
\caption{Structure of CRPFB.}
\label{fig2}
\end{figure*}
%\begin{figure}[!htbp]
%\setlength{\abovecaptionskip}{0pt}
%\setlength{\belowcaptionskip}{0pt}
%\centering\includegraphics[width=3.5in,angle=0]{fig2.pdf}
%\caption{Structure of CRPFB.}
%\label{fig2}
%\end{figure}

It can be seen from Fig.~2 that, besides input, original prior information and output, CRPFB includes two parts, denoting by blue rectangle and yellow rectangle respectively. The blue rectangle is parallel part, consisting of many parallel CPRFs, for example, $M$ CRPFs, using different and precise prior information determined by Equations (\ref{eq7}), (\ref{eq8}), or (\ref{eq9}).
The yellow rectangle is global part, meaning to find the minimum cumulated cost $C_{\text{cum}}^{m_{\text{min}}}$ among $C_{\text{cum}}^{1}, C_{\text{cum}}^{2}, \cdots, C_{\text{cum}}^{M}$. Then the estimated results $\hat{\textbf{X}}_{m_\text{min}}$ of the CRPF with this $C_{\text{cum}}^{m_{\text{min}}}$ is taken as the filtering results of CRPFB.

\subsection{Pseudo-code of CRPFB}
Taking the $m$-th CRPF as an exmple, the pseudo-code of CRPFB is presented as follows.
\begin{enumerate}[(1)]
  \item \textit{Initialization}\\
At initial time instant $t_{0}$, the particle-cost set of the $m$-th CRPF of CRPFB is initialized as $\{\textbf{x}_{t_{0},m}^{i},C_{t_{0},m}^{i}=0\}_{i=1}^{N}$, where $\textbf{x}_{t_{0},m}^{i}=[x_{val};\dot{x}_{t_{0},m}^{i}; y_{val};\dot{y}_{t_{0},m}^{i}]$,
$x_{val}\sim U[x_\text{min},x_\text{max}]$, $\dot{x}_{t_{0},m}^{i}$ is sampled from the suitable range in Equations~(\ref{eq7}), (\ref{eq8}), or (\ref{eq9}), the similar process can be used to initialize $y_{t_{0},m}$ and $\dot{y}_{t_{0},m}$. $C_{t_{0},m}^{i}$ is the cost of the particle $\textbf{x}_{t_{0},m}^{i}$, and $N$ is the number of particles used in the $m$-th CRPF.

  \item \textit{Resampling}\\
At time instant $t_{k}$, resample particle-cost set $\{\tilde{\textbf{x}}_{t_{k-1},m}^{j}, \tilde{C}_{t_{k-1},m}^{j}\}_{j=1}^{N}$ from updated particle-cost set $\{\textbf{x}_{t_{k-1},m}^{i},C_{t_{k-1},m}^{i}\}_{i=1}^{N}$ using “weight” $\mu(C_{t_{k-1},m}^{i})$.
For CRPF, $\tilde{C}_{t_{k-1},m}^{j}=C_{t_{k-1},m}^{i}$ if and only if $\tilde{\textbf{x}}_{t_{k-1},m}^{j}=\textbf{\textit{x}}_{t_{k-1},m}^{i}$.
$\mu(C_{t_{k-1},m}^{i})$ is defined in Equation~(\ref{eq10}).
\begin{equation}\label{eq10}
\begin{aligned}
\mu(C_{t_{k-1},m}^{i})&=\frac{(C_{t_{k},m}^{i})^{-q}}{\sum_{g=1}^{N}(C_{t_{k},m}^{g})^{-q}},\\
\end{aligned}
\end{equation}
where $q$ is a positive integer and the choice of $q$ will be discussed later in simulation parts.

  \item \textit{Updating}\\
At time instant $t_{k}$, take updated set $\{\textbf{x}_{t_{k},m}^{j},C_{t_{k},m}^{j}\}_{j=1}^{N}$ based on the resampled particle-cost set $\{\tilde{\textbf{x}}_{t_{k-1},m}^{j},\tilde{C}_{t_{k-1},m}^{j}\}_{j=1}^{N}$.

\begin{equation}\label{eq11}
\begin{aligned}
&\textbf{x}_{t_{k},m}^{j} \sim \aleph (\textbf{A}\tilde{\textbf{x}}_{t_{k-1},m}^{j},\textbf{Bv}),\\
&C_{t_{k},m}^{j}=\tilde{C}_{t_{k-1},m}^{j}+\vartriangle C_{t_{k},m}^{j},\\
&\vartriangle C_{t_{k},m}^{j}=\|\textbf{z}_{t_{k-1}}\|^{2}-\langle \textbf{z}_{t_{k-1}},f_{y}(\textbf{x}_{t_{k},m}^{j})\rangle^{2}.
\end{aligned}
\end{equation}

where $\aleph(\textbf{A}\tilde{\textbf{x}}_{k-1,m}^{j},\textbf{Bv})$ is Gaussian distribution with mean $\textbf{A}\tilde{\textbf{x}}_{t_{k-1},m}^{j}$ and covariance $\textbf{Bv}$. $\vartriangle C_{t_{k},m}^{j}$ is the incremental cost of the $m$-th CRPF at time step $t_{k}$, which is usually defined by users to measure the mismatch between the true state/observation and the estimated state/observation. Here we give an example of $\vartriangle C_{t_{k},m}^{j}$, where $\langle \cdot,\cdot \rangle$ denotes inner product.

%% sigma 的确定
 %$\sigma=\frac{|f_{\textrm{max}}-f_{\textrm{min}}|}{2K}$, $\textit{\textbf{A}}$ and \textit{\textbf{E}} are defined in Equation~(\ref{eq3}).
  \item \textit{State estimation of the m}-th \textit{CRPF of the CRPFB}\\
The estimation results of the $m$-th CRPF are denoted as $\hat{\textbf{X}}_{m}=\{\hat{\textbf{x}}_{t_{1},m},\hat{\textbf{x}}_{t_{2},m},\ldots,\hat{\textbf{x}}_{t_{K},m}\}$, where
\begin{equation}\label{eq12}
\hat{\textbf{x}}_{t_{k},m}=\hat{\textbf{x}}_{t_{k},m}^{j_{\textrm{min}}},j_{\textrm{min}}=\min_{j}\{C_{t_{k},m}^{j}\}_{j=1}^{N}.
\end{equation}
  \item \textit{Cumulated cost of the m}-th \textit{CRPF}\\
The cumulated cost $C_{\textrm{cum}}^{m}$ of the $m$-th CRPF is defined in Equation~(\ref{eq13}).
\begin{equation}\label{eq13}
  C_{\textrm{cum}}^{m}=\sum_{k=1}^{K}\left(\vartriangle C_{t_{k},m}\right).
  %C_{\textrm{cum}}^{m}=\sum_{k=1}^{K}\left(\|\textit{\textbf{z}}_{k}\|_{2}^{2}-\frac{|\textit{\textbf{z}}_{k}\textbf{h}^{H}(\hat{\textit{\textbf{x}}}_{m,k})|^{2}}{\|\textbf{h}(\hat{\textit{\textbf{x}}}_{m,k})\|_{2}^{2}}\right).
\end{equation}
  \item \textit{Comparison of cumulated costs}\\
Compare cumulated costs of $M$ CRPFs and label the minimum one.
\begin{equation}\label{eqa}
  m_{\textrm{min}}=\min_{m}\{C_{\textrm{cum}}^{1},C_{\textrm{cum}}^{2},\cdots,C_{\textrm{cum}}^{M}\}.
\end{equation}
  \item \textit{Filtering results of CRPFB}\\
The filtering results $\hat{\textbf{X}}_{\textrm{CRPFB}}$ of CRPFB is the estimated results of the $m_{\textrm{min}}$-th CRPF, which has the minimum cumulated cost among $C_{\textrm{cum}}^{1},C_{\textrm{cum}}^{2},\cdots,C_{\textrm{cum}}^{M}$.
\begin{equation}\label{eqb}
\begin{aligned}
  &\hat{\textbf{X}}_{\textrm{CRPFB}}=\hat{\textbf{X}}_{m_{\textrm{min}}},\\
  &\hat{\textbf{X}}_{m_{\textrm{min}}}=\{\hat{\textbf{x}}_{m_{\textrm{min},t_{1}}},\cdots,\hat{\textbf{x}}_{m_{\textrm{min},t_{K}}}\}.
\end{aligned}
\end{equation}
\end{enumerate}

Because the basic principle of CRPF is to consider the particle (or sample) with the minimum cost as the state estimate, the estimated state sequence provided by the CRPF with the minimum cumulated cost should also be considered as the estimated results of CRPFB. In the CRPFB, all the CRPFs use the same parameters except prior information; hence it can be said that the CRPF with the minimum cumulated cost in the CRPFB uses the prior information that is much closer to the true state. Or we can say that the CRPF using much more precise prior information should provide the best state estimation results.

\subsection{TBD algorithm based on CRPFB}

The structure of the proposed CRPFB based batch TBD algorithm is shown in Fig.~3, which includes four steps:
\begin{enumerate}[(1)]
  \item estimate the state sequence of a possible signal by using CRPFB;
  \item calculate the test metric $\Psi(\textbf{Z}_{K},\hat{\textbf{X}}_{\textrm{CRPFB}})$ based on the state estimation provided by the CRPFB;
  \item compare $\Psi(\textbf{Z}_{K},\hat{\textbf{X}}_{\textrm{CRPFB}})$ with the given threshold $V_{T}$;
  \item declare signal presence if $\Psi(\textbf{Z}_{K},\hat{\textbf{X}}_{\textrm{CRPFB}})$ exceeds $V_{T}$, otherwise declare signal absence.
\end{enumerate}

In Fig.~3, the test metric $\Psi(\textbf{Z}_{K},\hat{\textbf{X}}_{\textrm{CRPFB}})$ is defined as follows.
\begin{equation}\label{eq16}
  \Psi(\textbf{Z}_{K},\hat{\textbf{X}}_{\textrm{CRPFB}})=\sum_{k=1}^{K}\langle \textbf{z}_{k},f_{y}(\hat{\textbf{x}}_{m_{\textrm{min},t_{k}}})\rangle^{2},
\end{equation}
where $\textbf{Z}_{K}=\{\textbf{z}_{t_{1}}, \textbf{z}_{t_{2}}, \cdots, \textbf{z}_{t_{K}}\}$.

Moreover, $\textbf{Z}_{K-\text{noise}, m_{c}}$ denotes the $m_{c}$-th noise-only observation,
$\Psi_{m_{c}}(\textbf{Z}_{K},\hat{\textbf{X}}_{\textrm{CRPFB}})$ is the test metric computed from the $m_{c}$-th noise-only observation, and $M_{c}$ is the total number of noise-only observations used to estimate the threshold $V_{T}$ under the given false alarm probability $P_{\text{fa}}$.

\begin{figure*}[!t]
\setlength{\abovecaptionskip}{0pt}
\setlength{\belowcaptionskip}{0pt}
\centering
\includegraphics[width=5in,angle=0]{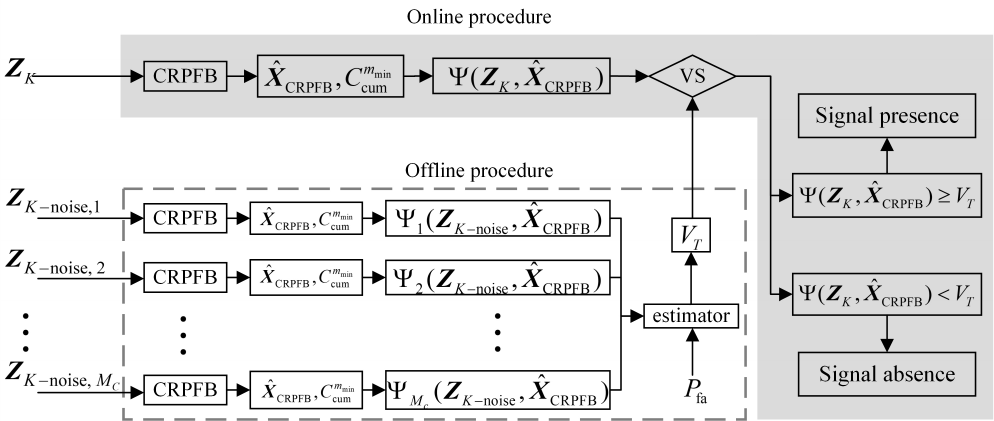}
\caption{Structure of CRPFB based detector.}
\label{fig3}
\end{figure*}

It can be seen from Fig.~3 that the proposed batch TBD algorithm includes an online procedure and an offline procedure.
The online procedure is the shaded part to determine the presence or absence of a target by comparing the test metric with a given threshold $V_{T}$. A target is declared to be present if $\Psi(\textbf{Z}_{K},\hat{\textbf{X}}_{\text{CRPFB}})$
exceeds the given threshold $V_{T}$, or no target is declared.

In Fig.~3, the offline procedure in the dashed box is to estimate the threshold $V_{T}$ under the given false alarm probability $P_{\text{fa}}$, which can be estimated from the test metrics
$\Psi_{1}(\textbf{Z}_{K-\text{noise}}, \hat{\textbf{X}}_{\text{CRPFB}}), \cdots, \Psi_{M_{c}}(\textbf{Z}_{K-\text{noise}}, \hat{\textbf{X}}_{\text{CRPFB}})$ of the noise-only observations. Traditional numerical methods such as ranking is available, but requires a large number of noise-only observations.
In next subsection, an analytical method for estimating the threshold $P_{\text{fa}}$ is introduced, which needs much less noise-only observations.

\subsection{Analytical threshold estimation of test statistic for batch detection}
Generally, the test metrics of the PF based detection algorithms have unknown probability distributions, because the probability distributions of state estimation is always unknown.
Usually, numerical method is used to estimate the threshold in this case.
In numerical method, the test metric calculated from noise-only observations are sorted in an ascending order and the test metric related to $1-P_{fa}$ is considered as the threshold. This method is simple and requires no statistical information on the test metrics. However, a large number of observations are required to obtain an accurate detection threshold. For example, in the case of $P_{fa}=10^{-3}$, nearly $4\times10^{7}$ noise-only observations are required to estimate the threshold with 95\% confidence and 1\% relative error~\cite{van2004detection}.

By contrast, if the distribution of the test metric of the detector was known, the threshold estimate can be easier.
For our proposed method, it can be seen from Fig.~2 and Fig.~3 that the test metric $\Psi_{m_{c}}(\textbf{Z}_{t_{K}},\hat{\textbf{X}}_{\textrm{CRPFB}})$ is computed from the maxima cumulated cost among all the CRPFs of the CRPFB and thus it obeys GEV distribution \cite{b20,kotz2000extreme}. Next, $\Psi_{m_{c}}(\textbf{Z}_{t_{K}},\hat{\textbf{X}}_{\textrm{CRPFB}})$ is briefly denoted as $\Psi_{m_{c}}$.

The GEV distribution function $G(\Psi_{m_{c}}|\kappa,\rho,\eta)$ with location parameter $\rho$, scale parameter $\eta$, and shape parameter $\kappa$ is shown in Equation~(\ref{eq19}), with $-\infty<\rho<\infty, -\infty<\kappa<\infty,$ and $\eta>0$. The maximum likelihood estimates of the three parameters can be obtained from fewer test metrics in the case of noise-only observations.
In this paper, we use the function "gevfit" in MATLAB2016B to estimate the three parameters of GEV distribution.
Based on the GEV distribution function, the threshold $V_{T}$ with given false alarm $P_{fa}$ can be estimated as illustrated in Equation~(\ref{eq21}).
\begin{equation}\label{eq19}
\begin{aligned}
G(\Psi_{m_{c}}|\kappa,\rho,\eta)=\textrm{exp} \left\{-\left[1+\kappa\left(\frac{\Psi_{m_{c}}-\rho}{\eta}\right)\right]\right\},
\end{aligned}
\end{equation}
where $1+\kappa\frac{\Psi_{m_{c}}-\rho}{\eta}>0$.
\begin{equation}\label{eq21}
V_{T}=G^{-1}(1-P_{fa}).
\end{equation}

\section{Numerical experiments}
In this section, we use two types of test signals to evaluate the superior performance of the proposed method on detection capability, estimation accuracy, and computational complexity. Moreover, we show that the test metric on the CRPFB based detector fits the GEV distribution and its detection threshold can be analytically estimated from fewer noise-only measurements using the GEV theory.
\subsection{Two types of test signals}
In this subsection, the two types of test signals we used are shown and the first one is first used in \cite{shui2016detection}. The first test signal is as follows.
\begin{equation}\label{eq24}
\begin{aligned}
z_{1}(t)&=s_{1}(t)+w(t)\\
&=a(1+b\cos(12\pi t))\cdot\\
&\text{exp}\left(2\pi j\left(a_{1}t+\frac{a_{2}t^{2}}{2}+\frac{a_{3}t^{3}}{3}+\frac{a_{4}t^{4}}{4}\right)\right),\\
\end{aligned}
\end{equation}
where $a(1+b\cos(12\pi t))$ denotes the time-varying amplitude which is determined by its SNR, $b\in[0,1]$, $a_{1}$, $a_{2}$, $a_{3}$, $a_{4}\in[-20,20]$, and $t\in[0,T]~ \textrm{s}$. The measurement noise $w(t)$ obeys a complex generalized Gaussian distribution with variance 1 and shape parameter 0.5 \cite{novey2009complex}, and it is assumed to be unknown. The frequency, chirp rate and SNR of $s_{1}(t)$ are as follows.
\begin{equation}\label{eq25}
\begin{aligned}
&f_{1}(t)=a_{1}+a_{2}t+a_{3}t^{2}+a_{4}t^{3},\\
&f_{1}^{\prime}(t)=a_{2}+2a_{3}t+3a_{4}t^{2},\\
&\text{SNR}=10~\textrm{log}\left(a^{2}\left(1+\frac{b^{2}}{2}\right)\right)(\textrm{dB}).
\end{aligned}
\end{equation}

The second test signal is as follows.
\begin{equation}
\begin{aligned}
z_{2}(t)&=s_{2}(t)+w(t)\\
&=a~\textrm{exp}(-jb~\textrm{cos}(2\pi t))+w(t).\\
\end{aligned}
\end{equation}
where $a$ is determined by its SNR, $b\in[-40,40]$, and $t\in[0,T]~\textrm{s}$. The frequency, chirp rate and SNR of $s_{2}(t)$ are as follows.
\begin{equation}\begin{aligned}
&f_{2}(t)=b~\textrm{sin}(2\pi t),\\
&f_{2}^{\prime}(t)=2\pi b~\textrm{cos}(2\pi t),\\
&\text{SNR}=20~\textrm{log}(a)(\textrm{dB}).
\end{aligned}\end{equation}

\subsection{Performance of CRPFB based detector}
In this subsection, the detection performance of the proposed CRPFB based detector (Proposed) are compared with the CRPF-detection method (CRPF-detector) \cite{b12}, FB-CRPF based detection method (FB-CRPF-detector) \cite{shui2016detection} in terms of detection probability curve, receiver operating characteristic (ROC) curve, and execution time.
Moreover, the impacts of amplitude fluctuation and subinterval length on the detection performance of the CRPFB-detector are discussed.
Finally, the reason for the superior performance of the proposed method is explored. For all the three methods, the possible signal is approximated as a piecewise LFM signal and they use the same cost function as shown in Equation (\ref{eq11}).
%\begin{figure}[!h]
%\centering
%\subfigure{
%\begin{minipage}[!c]{0.5\textwidth}
%%\centering
%    \includegraphics[width=3.5in]{fig4a.jpg}
%\end{minipage}
%}
%\subfigure{
%\begin{minipage}[!c]{0.5\textwidth}
%%\centering
%    \includegraphics[width=3.5in]{fig4b.jpg}
%\end{minipage}
%}
%\subfigure{
%\begin{minipage}[!c]{0.5\textwidth}
%%\centering
%    \includegraphics[width=3.5in]{fig4c.jpg}
%\end{minipage}
%}
%\caption{Detection probabilities of the test signals provided by CRPF-detector, FB-CRPF-detector, CCM-detector and the proposed method. (a) Detection probabilities of $s_{1}(t)$, $T=0.5s$, $b=0.6$; (b) Detection probabilities of $s_{1}(t)$, $T=1s$, $b=0.6$; (c) Detection probabilities of $s_{2}(t)$, $T=1s$.}
%\label{fig4}
%\end{figure}

\begin{figure*}[!t]
\setlength{\abovecaptionskip}{0pt}
\setlength{\belowcaptionskip}{0pt}
\centering
\includegraphics[width=6in,angle=0]{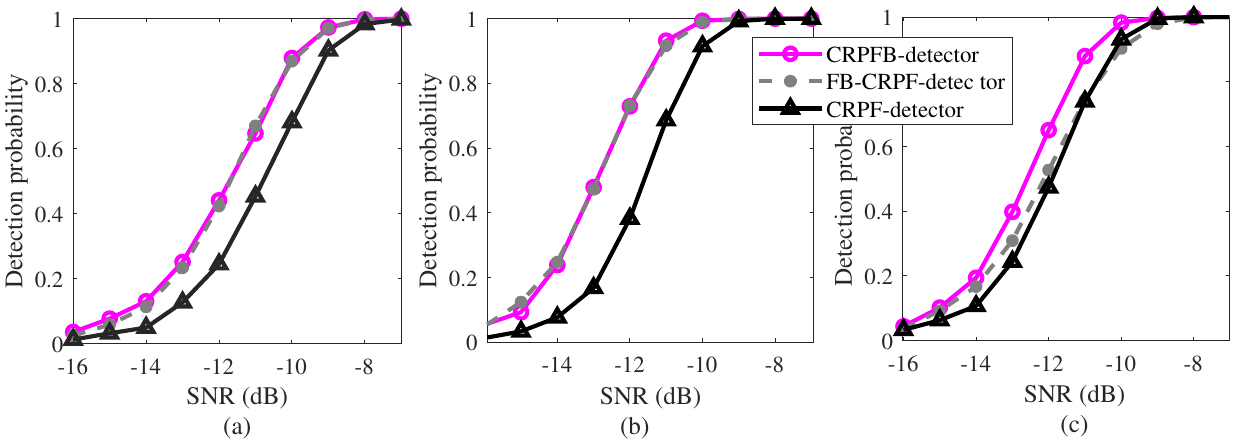}
\caption{Detection probabilities of the test signals provided by the CRPF-detector, the FB-CRPF-detector, the CCM-detector, and the proposed method (CRPFB-detector). (a) Detection probabilities of $s_{1}(t)$, $T=0.5~\textrm{s}$, $b=0.6$, (b) Detection probabilities of $s_{1}(t)$, $T=1~\textrm{s}$, $b=0.6$, (c) Detection probabilities of $s_{2}(t)$, $T=1~\textrm{s}$.}
\label{fig4}
\end{figure*}

Fig.~4 shows the detection probability curves provided by the three methods, where SNR varies from -16dB to -7dB. In Fig.~4, the sampling time is $t_s=\frac{1}{512}~\textrm{s}$, $\triangle T=\frac{1}{16}~\textrm{s}$, and the probability of false alarm is $P_{\text{fa}}=10^{-3}$.
The thresholds for the four methods are estimated by numerical method using $10^{5}$ measurements with noises only.
The detection probabilities of the four methods are estimated by using 2,000 noisy signals.

In Fig.~4(a), the test signal is $s_{1}(t)$, $t\in[0,0.5]~\textrm{s}$, $b=0.6$, which meets the condition in Equation (\ref{eq7}).
In Fig.~4(b), the test signal is $s_{1}(t)$, $t\in[0,1]~\textrm{s}$, $b=0.6$, which meets the condition in Equation (\ref{eq8}).
In Fig.~4(c), the test signal is $s_{2}(t)$, $t\in[0,1]~\textrm{s}$, which meets the condition in Equation (\ref{eq9}).

CRPF-detector uses 200, 400 and 800 particles respectively in Fig.~4(a)-Fig.~4(c), so as FB-CRPF-detector.
For the proposed CRPFB-detectors in Fig.~4(a) and Fig.~4(b), the proposed CRPFB-detector consists of 2000 CRPFs and 1 particle is used in each CRPF.
For the proposed CRPFB-detector in Fig.~4(c), the measurement is divided into 4 blocks $[0,\frac{1}{4}]~\textrm{s}$, $[\frac{1}{4},\frac{1}{2}]~\textrm{s}$, $[\frac{1}{2}, \frac{3}{4}]~\textrm{s}$, and $[\frac{3}{4}, 1]~\textrm{s}$ .
For each block, the measurement is divided into $K=4$ equi-length subinterval and a CRPFB is used. The number of CRPFs used in each CRPFB is 2,000 and the number of particles used in each CRPF is 1.
For all the results in Fig.~4, the parameter $q$ in Equation~(\ref{eq10}) is 5.

Besides detection probabilities, the ROC curves of the three methods in Fig.~5 are also used to illustrate the detection capability of the proposed method with SNR=-11dB. It can be seen from Fig~4 and Fig.~5 that the CRPFB based detector and the FB-CRPF-detector performs better in detection.
\begin{figure*}[!h]
\setlength{\abovecaptionskip}{0pt}
\setlength{\belowcaptionskip}{0pt}
\centering
\includegraphics[width=6in,angle=0]{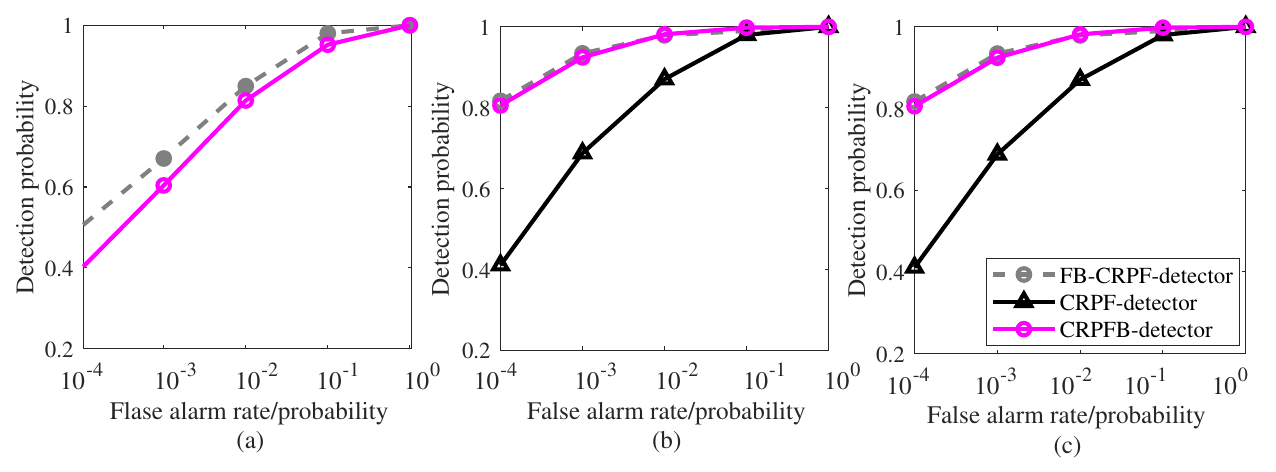}
\caption{ROC curve of the test signals provided by CRPF-detector, FB-CRPF-detector, and the proposed method (CRPFB-detector). (a) Detection probabilities of $s_{1}(t)$, $T=0.5~\textrm{s}$, $b=0.6$, (b) Detection probabilities of $s_{1}(t)$, $T=1~\textrm{s}$, $b=0.6$, (c) Detection probabilities of $s_{2}(t)$, $T=1~\textrm{s}$.}
\label{fig5}
\end{figure*}

\begin{figure*}[!h]
\setlength{\abovecaptionskip}{0pt}
\setlength{\belowcaptionskip}{0pt}
\centering
\includegraphics[width=6in,angle=0]{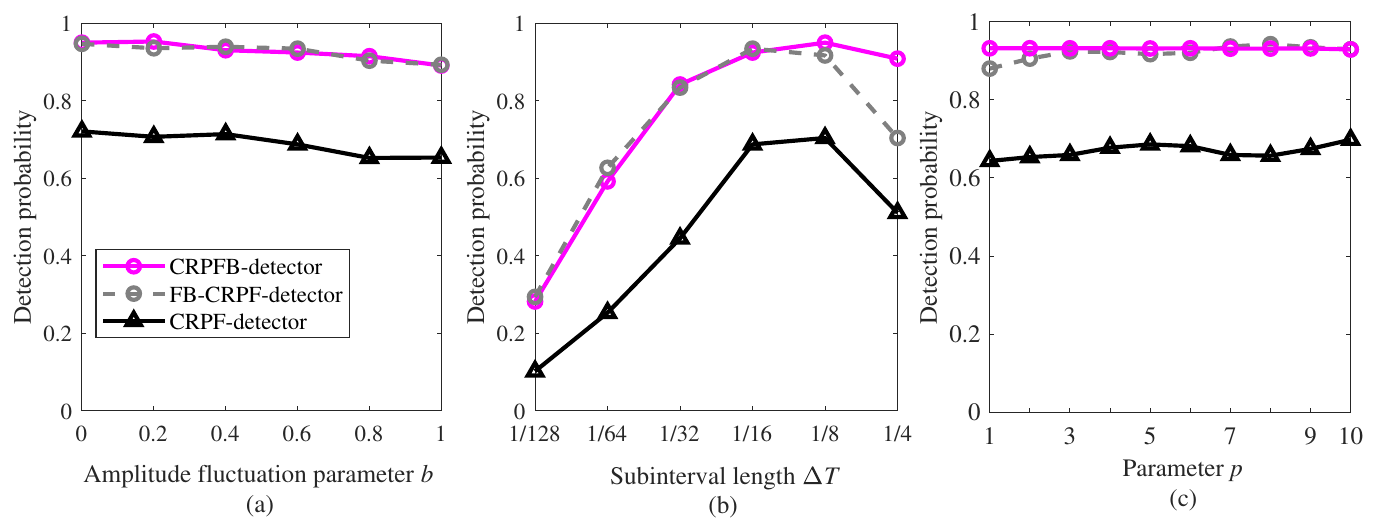}
\caption{Impacts of amplitude fluctuation, subinterval length and the parameter $q$ on detection performance of CRPFB-detector, CRPF-detector, and FB-CRPF-detector, SNR=-11dB. (a) Detection probabilities versus $b$, $\triangle T=\frac{1}{16}~\textrm{s}$, $q=5$. (b) Detection probabilities versus $\triangle T$, $b=0.6, q=5$. (c) Detection probabilities versus $q$, $b=0.6$, $\triangle T=\frac{1}{16}~\textrm{s}$.}
\label{fig6}
\end{figure*}

The execution times of the three methods for the three simulations are summarized in Table I, where the execution time is the average time cost of a method to detect a test signal. For the proposed method, because of its parallel architecture, the execution time is the summation of the estimation time of one CRPF of the CRPFB and the detection time. Table I indicates that the execution time of the proposed method is far shorter than that of the other methods.
\begin{table}[!htbp]
\caption{Execution time of the four detection methods for the three simulations, SNR=-11dB, $P_{\text{fa}}=10^{-3}$.}
\centering
\label{table}
\small
\setlength{\tabcolsep}{3pt}
\begin{tabular}{|c|c|c|c|}
%{|p{46pt}|p{31pt}|p{41pt}|p{40pt}|p{40pt}|}
\hline
\makecell[c]{Methods}& \makecell[c]{Simulation 1 (s)} &\makecell[c]{Simulation 2 (s)} &\makecell[c]{Simulation 3 (s)}\\
\hline
CRPF-detector&$0.013$&$0.052$&$0.104$ \\
\hline
\makecell[c]{FB-CRPF\\detector}&$0.025$&$0.102$ &$0.202$\\
\hline
Proposed&$1.414\times10^{-4}$&$2.445\times10^{-4}$&$2.690\times10^{-4}$ \\
\hline
\end{tabular}
\label{tab1}
\end{table}

Next, we explore the impacts of several parameters on CRPF-detector, FB-CRPF-detector, and the proposed CRPFB-detector. In Fig.~6, the test signal is $s_{1}(t), T\in[0,1]~\text{s}$, which satisfies the condition in Equation (\ref{eq8}). Moreover, the numbers of particles used by CRPF-detector and FB-CRPF-detector, the number of CRPFs in the proposed method and the number of particles in each CRPF of CRPFB-detector are the same as that in Fig.~5(b).
Fig.~6 shows the detection performances of the three methods varying with amplitude fluctuation, subinterval length and the parameter $q$ in Equation (\ref{eq10}).
In Fig.~6(a), $b=0.1,0.2,\cdots,0.9$ controls the amplitude fluctuation of $s_{1}(t)$ and large $b$ represents strong fluctuation, $\triangle T=\frac{1}{16}$, $q=5$;
in Fig.~6(b), subinterval length $\triangle T=\frac{1}{128}, \frac{1}{64}, \frac{1}{32}, \frac{1}{16}, \frac{1}{8}, \frac{1}{4}$, $b=0.6$, $q=5$;
in Fig.~6(c), the parameter in Equation (\ref{eq10}) is set as $q=1, 2, \cdots, 10$, $b=0.6$, $\triangle T=\frac{1}{16}$.

Fig.~6(a) shows that the impact of amplitude fluctuation on the proposed method and other two methods is small, because the amplitude of possible signal in  the state-space model we used is removed.
Fig.~6(b) indicates that the three methods achieve best detection performances with $\triangle T=\frac{1}{8}$, which means that a proper subinterval length is necessary and the piecewise LFM model is suitable with this subinterval length.
Fig.~6(c) reveals that the parameter $q$ has no impact on the proposed CRPFB-detector, because this parameter is related with resampling process and the resampling process is removed in the proposed method with only 1 particle using in each CRPF of CRPFB. On the other hand, the parameter $q$ has slightly impact on the other two methods. CRPF-detector achieves best performance with $q=5$, and FB-CRPF-detector obtains good performance when $q\geq 3$.

The aforementioned simulation results illustrate that, the detection probability of the proposed method is better than that of CRPF-detector in all the three simulations. Compared with FB-CRPF-detector, the detection probability of the proposed method is comparable in the first two simulations and is better in the third simulation. Moreover, the execution time of the proposed method is much shorter than that of CRPF-detector and FB-CRPF-detector.
For example, the execution time of the proposed method is about 1 percent of that of FB-CRPF-detector.
In addition, the impacts of amplitude fluctuation and parameter $q$ in Equation (\ref{eq10}) on the proposed method are small, and thus the proposed method is more robust.

\subsection{Results analysis of CRPFB based detector}
In this part we explore why the proposed method has good detection performance among the three methods. For CRPF-detector, the FB-CRPF-detector, and the proposed method, the test metrics are GLR, bi-feature including GLR and TV, and GLR.
These test metrics are obtained by using the IF estimations. Therefore, the accuracy of IF estimation is crucial for detection capability of these methods, and the execution time of IF estimation is crucial for detection speed.
Fig.~7(a), (b), and (c) shows the root mean square errors (RMSEs) of the IF estimations of the test signals corresponding to Fig.~4(a), (b), and (c) respectively, where the RMSE is calculated in terms of Equation~(\ref{eq26}),
\begin{equation}\label{eq26}
  \textrm{RMSE}=\sqrt{\frac{T}{t_{s}}\sum_{l_{s}=1}^{L_{s}}(f-\hat{f})}.
\end{equation}
where $\hat{f}$ is the estimation of the IF curve, with $\hat{f}=[\hat{f}_{1},\hat{f}_{2},\ldots, \hat{f}_{K}]$, $\hat{f}_{k}=\hat{x}_{k}(1)+\hat{x}_{k}(2)t_{s}\textit{\textbf{L}}$, $\textit{\textbf{L}}=0,1,\ldots, 31$, and $L_{s}=T/t_{s}$. Table II shows the execution times of the CRPF, FB-CRPF, and CRPFB to estimate the IF curve of test signals, where the execution time is to estimate the IF curve of a test signal. It can be seen from Fig.~7 and Table II that our CRPFB method achieves the best accuracy of IF estimation using the shortest execution time. Thus, the CRPFB is the main reason for the good detection performance of the proposed method.

\begin{figure*}[!t]
\setlength{\abovecaptionskip}{0pt}
\setlength{\belowcaptionskip}{0pt}
\centering
\includegraphics[width=6in,angle=0]{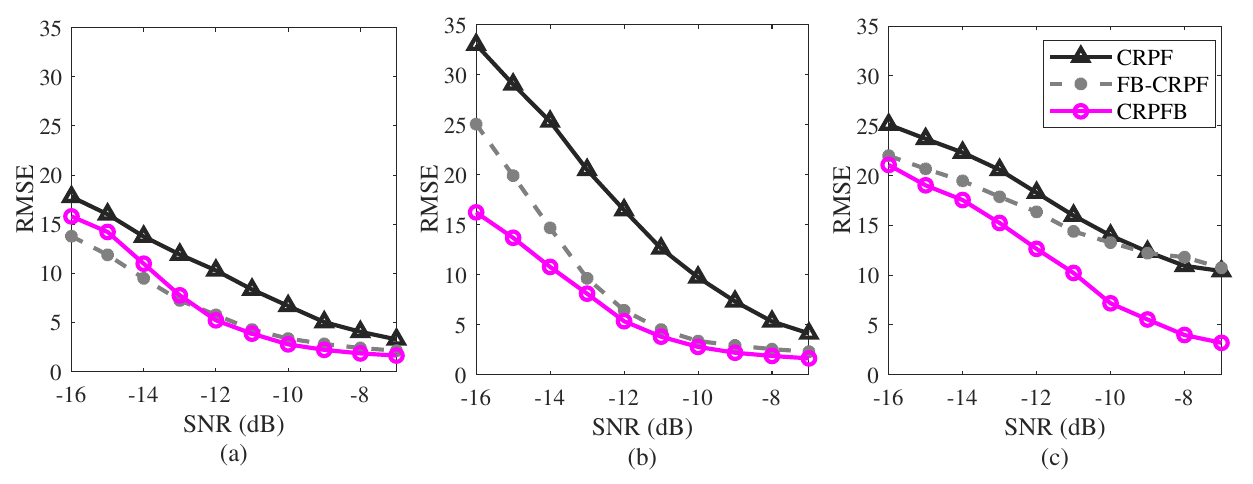}
\caption{RMSEs of the test signals provided by CRPF, FB-CRPF, CCM, and the proposed CRPFB, (a) $s_{1}(t)$, $T=0.5~\textrm{s}$, $b=0.6$, (b) $s_{1}(t)$, $T=1~\textrm{s}$, $b=0.6$, (c) $s_{2}(t)$, $T=1~\textrm{s}$.}
\label{fig7}
\end{figure*}

\begin{table}[!htbp]
\caption{Execution time of the four estimation methods for the three simulations, SNR=-11dB, $P_{fa}=10^{-3}$.}
\centering
\label{table}
\small
\setlength{\tabcolsep}{3pt}
\begin{tabular}{|c|c|c|c|}
%{|p{46pt}|p{31pt}|p{41pt}|p{40pt}|p{40pt}|}
\hline
\makecell[c]{Methods}& \makecell[c]{Simulation 1 (s)} &\makecell[c]{Simulation 2 (s)} &\makecell[c]{Simulation 3 (s)}\\
\hline
CRPF&$0.012$&$0.050$&$0.103$ \\
\hline
FB-CRPF&$0.024$&$0.10$ &$0.200$\\
\hline
CRPFB&$1.412\times10^{-4}$&$2.440\times10^{-4}$&$2.680\times10^{-4}$ \\
\hline
\end{tabular}
\label{tab2}
\end{table}

\subsection{Performance of the CRPFB}
In this subsection we discuss the performance of the CRPFB, because it is the key factor for the good detection performance and short execution time of the proposed method.

The simulation results in Section VB indicate that the CRPFB based detector obtains good detection performance with short execution time.
The reason is that the proposed CRPFB has an entirely parallel architecture and uses precise prior information.
Actually, there have been many attempts for parallel architectures for PFs,
which mainly focus on parallel architectures of resampling procedures for PFs.
These resampling architectures for parallel implementation mainly include local resamplling (LR) \cite{miguez2007analysis}, resampling with nonpropotional allocation (RNA) \cite{bolic2005resampling}, and resampling using Metropolis Hastings (IMH) sampler \cite{sankaranarayanan2008algorithmic}.

\begin{figure*}[!t]
\setlength{\abovecaptionskip}{0pt}
\setlength{\belowcaptionskip}{0pt}
\centering
\includegraphics[width=6in,angle=0]{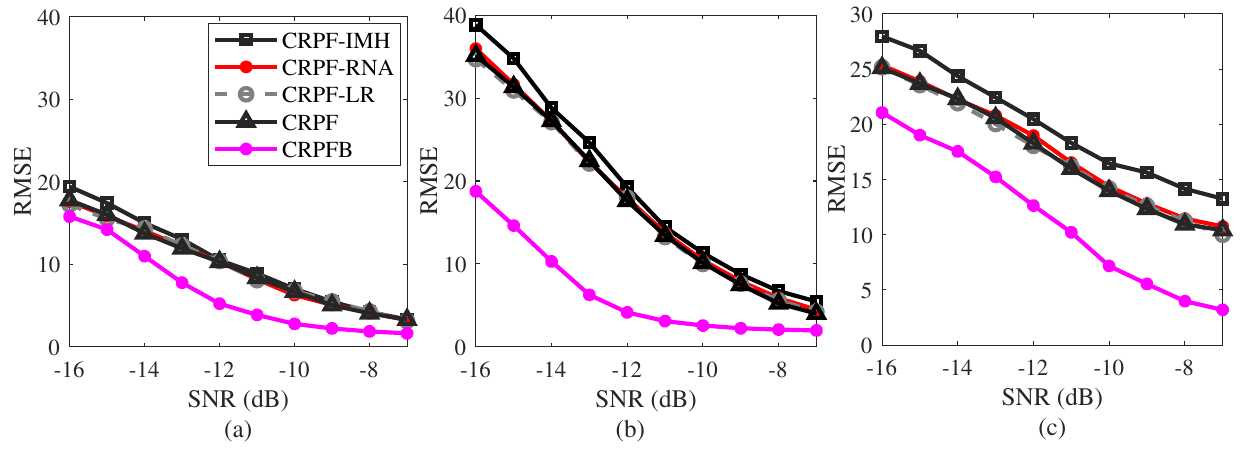}
\caption{RMSEs of the two test signals provided by CRPF, CRPF using LS, RNA, IMH, and CRPFB,(a) $s_{1}(t)$, $T=0.5~\textrm{s}$, $b=0.6$, (b) $s_{1}(t)$, $T=1~\textrm{s}$, $b=0.6$, (c) $s_{2}(t)$, $T=1~\textrm{s}$.}
\label{fig8}
\end{figure*}

\begin{figure*}[!t]
\setlength{\abovecaptionskip}{0pt}
\setlength{\belowcaptionskip}{0pt}
\centering
\includegraphics[width=6in,angle=0]{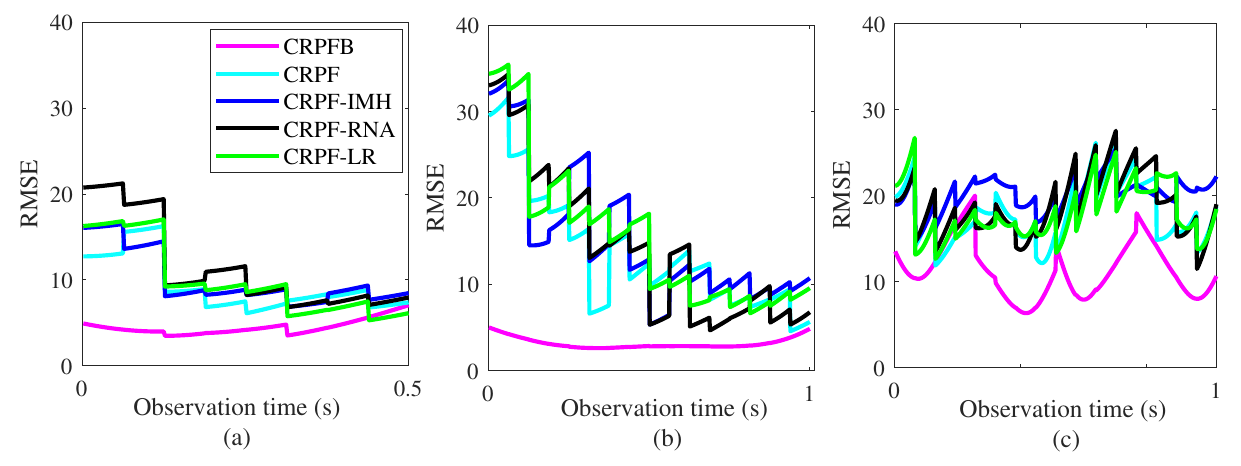}
\caption{RMSEs of the three types of test signals vary with observation time provided by CRPF, CRPF using LS, RNA, IMH, and CRPFB, SNR=-11dB, (a) $s_{1}(t)$, $T=0.5~\textrm{s}$, $b=0.6$, (b) $s_{1}(t)$, $T=1~\textrm{s}$, $b=0.6$, (c) $s_{2}(t)$, $T=1~\textrm{s}$.}
\label{fig9}
\end{figure*}

\begin{figure*}[!t]
\setlength{\abovecaptionskip}{0pt}
\setlength{\belowcaptionskip}{0pt}
\centering
\includegraphics[width=6in,angle=0]{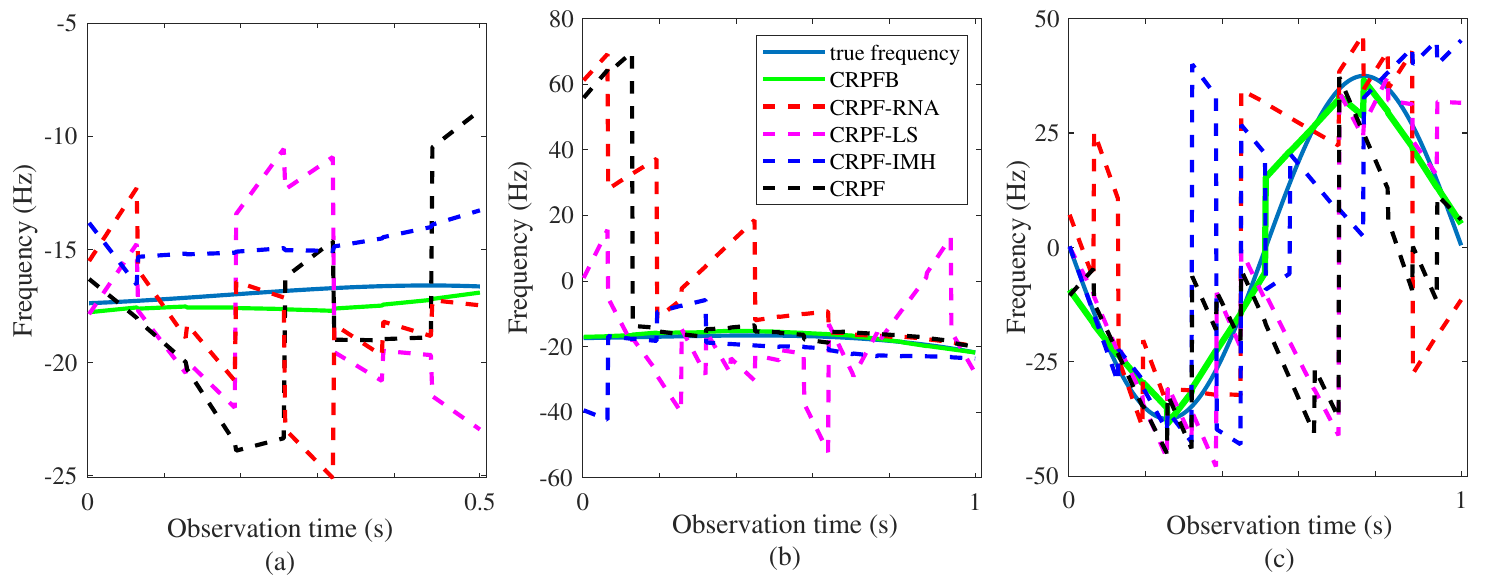}
\caption{True frequencies versus estimated IFs provided by CRPF, CRPF using LS, RNA, IMH, and CRPFB, SNR=-11dB, (a) $s_{1}(t)$, $T=0.5~\textrm{s}$, $b=0.6$, (b) $s_{1}(t)$, $T=1~\textrm{s}$, $b=0.6$, (c) $s_{2}(t)$, $T=1~\textrm{s}$.}
\label{fig14}
\end{figure*}

Fig.~8 shows the RMSEs versus SNRs, provided by the CRPFB, CRPF, CRPF using LR (CRPF-LR), CRPF using IMH (CRPF-IMH), and CRPF using RNA (CRPF-RNA). Fig.~9 shows the corresponding RMSEs at each time instant. Fig.~10 shows the estimated IF of the three kinds of test signals overlaid on true frequencies (random examples of the test signals). The test signals in Fig.~8, Fig.~9, and Fig.~10 are the same as that in Fig.~4.

The parameters used in Fig.~8(a), Fig.~9(a), and Fig.~10(a) are as follows:
the CRPF-RNA uses 200 particles and 50 processing elements;
the CRPF-LR uses 200 particles and 200 processing elements;
the CRPF-IMH uses 200 particles and 100 independent Metropolis Hastings samplers;
and the CRPFB uses 2,000 CRPFs and 1 particle in each CRPF.

The parameters used in Fig.~8(b), Fig.~9(b), and Fig.~10(b) are as follows:
the CRPF-RNA uses 400 particles and 100 processing elements;
the CRPF-LR uses 400 particles and 400 processing elements;
the CRPF-IMH uses 400 particles and 200 independent Metropolis Hastings samplers;
and the CRPFB uses 2,000 CRPFs and 1 particle in each CRPF.

The parameters used in Fig.~8(c), Fig.~9(c), and Fig.~10(a) are as follows.
the CRPF-RNA uses 800 particles and 200 processing elements;
the CRPF-LR uses 800 particles and 800 processing elements;
the CRPF-IMH uses 800 particles and 400 independent Metropolis Hastings samplers;
and the CRPFB uses 2,000 CRPFs and 1 particle in each CRPF.

It can be seen from Fig.~8, Fig.~9, and Fig.~10 that the CRPFB has the best estimation accuracy and the fastest convergence.

Fig.~11 shows the execution times for the aforementioned methods to process $s_{1}(t)$ with $t\in[0,0.5]~\textrm{s}$, $s_{1}(t)$ with $t\in[0,1]~\textrm{s}$, and $s_{2}(t)$ with $t\in[0,1]~\textrm{s}$, where the execution time of each method is the time of each method to estimate the IF curve of a test signal. For these CRPFs using parallel resampling algorithms, the execution time is approximated as the total time of estimating a test signal divided by the number of parallel resamplers. It can be seen that the execution time of the CRPFs using parallel sampling procedures is greatly shortened. And CRPF-LR and CRPFB have the shortest execution time.

\begin{figure*}
  \centering
  \includegraphics[width=3in]{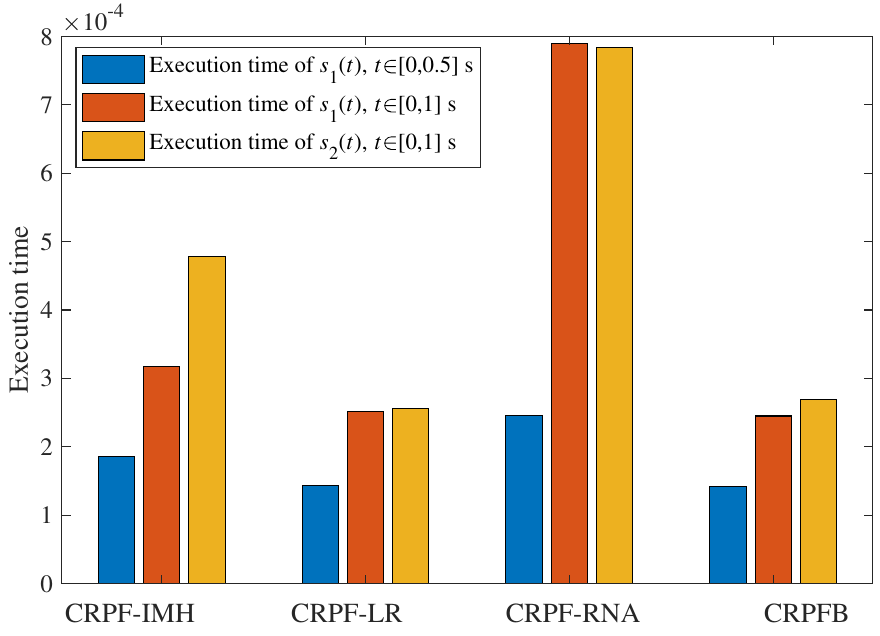}
  \caption{Execution time comparison of CRPF-IMH, CRPF-LR, CRPF-RNA, and CRPFB.}\label{fig10}
\end{figure*}

The simulation results in Figs.~8, 9, 10 and 11 show that, compared with the CRPFs using parallel resampling algorithms, the CRPFB achieves  best IF estimation accuracy with shortest execution time.

It is worth to note that there are two important parameters in the CRPFB, the number of CRPFs used in the CRPFB and the number of particles used in each CRPF. Next we explore the impact of these two parameters on the IF estimation accuracy of the CRPFB. Fig.~12 shows that the RMSEs of $s_{1}(t)$, $t\in[0,0.5]~\textrm{s}$, $s_{1}(t)$, $t\in[0,1]~\textrm{s}$, and $s_{2}(t)$, $t\in[0,1]~\textrm{s}$ vary with the number of particles used in each CRPF and the number of CRPFs used in the CRPFB.
It can be seen that the proposed CRPFB obtains best estimation results when the number of particles used in CRPFs of CRPFB is 1. And the RMSEs of the test signals increase with the increases of the CRPFs. The reason is that the prior information provided for the CRPFs of the CRPFB is more precise with the increases in the number of CRPFs.
Moreover, the principle of determining the number of CRPFs of CRPFB is unclear. It can be seen from Fig.~12 that, with 1 particle used in each CRPF, the proposed CRPFB obtains similar estimation results when the number of CRPFs of CRPFB is larger than 1000.

\begin{figure*}[!t]
\setlength{\abovecaptionskip}{0pt}
\setlength{\belowcaptionskip}{0pt}
\centering
\includegraphics[width=6in,angle=0]{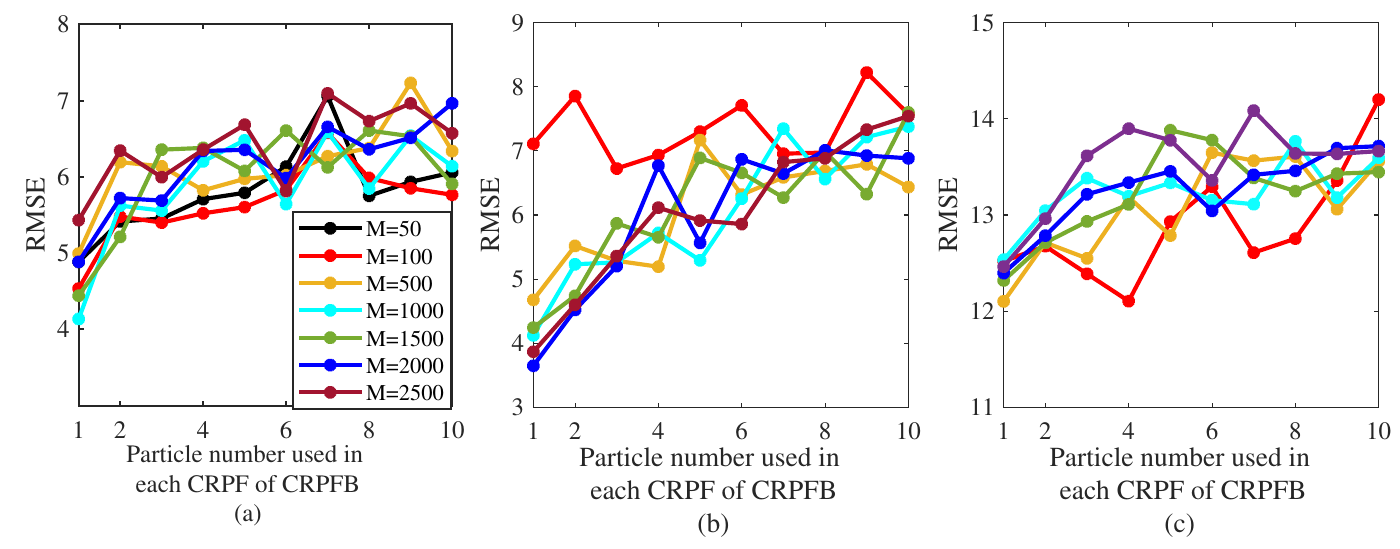}
\caption{RMSEs of the two test signals provided by the CRPFB using different numbers of particles and CRPFs, SNR=-10dB, (a) Detection probabilities of $s_{1}(t)$, $T=0.5~\textrm{s}$, $b=0.6$, (b) Detection probabilities of $s_{1}(t)$, $T=1~\textrm{s}$, $b=0.6$, (c) Detection probabilities of $s_{2}(t)$, $T=1~\textrm{s}$.}
\label{fig11}
\end{figure*}

It is an important issue to determine the number of CRPFs used in the CRPFB. Simulation results in Fig.~12 shows that the number of particles required by each CRPF can be reduced to 1 with a large enough number of CRPFs. And the CRPFB can be fully and quickly implemented in parallel. However, it is hard to determine the proper number of CRPFs for a special problem. The problem is similar to the problem of determining the number of particles in PFs. However, the simulation results shown from Fig.~5 to Fig.~12 indicate that, for a special problem, the number of CRPFs required by the CRPFB is nearly 4$\sim$5 times of the number of particles required by PFs.

\subsection{Evaluation of detection threshold $V_{T}$ via generalized extreme value theory}
In this subsection, first, we show that the proposed method under hypothesis $H_{0}$ fits the GEV distribution. Then the detection threshold under a given false alarm $P_{fa}$ is estimated by using the GEV distribution.

\begin{figure*}[!h]
\setlength{\abovecaptionskip}{0pt}
\setlength{\belowcaptionskip}{0pt}
\centering
\includegraphics[width=6in,angle=0]{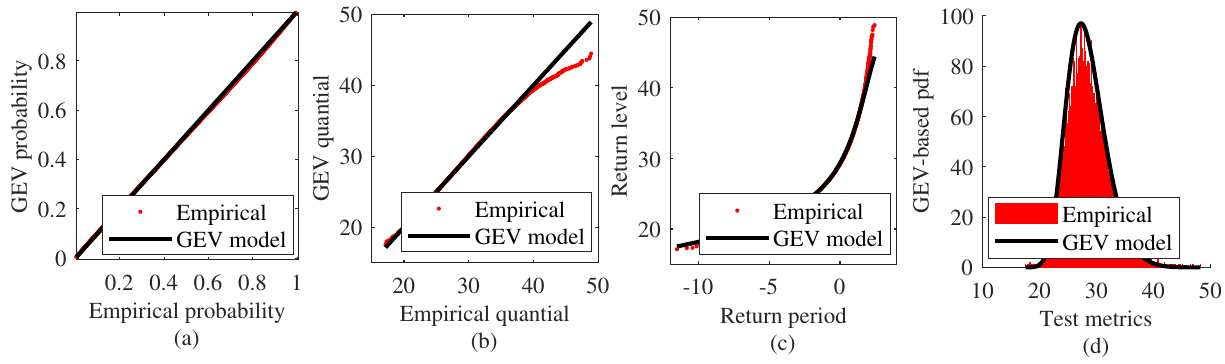}
\caption{Diagnostic plots for GEV fit to the test metrics of the CRPFB based detector, (a) Probability plot, (b) Quantile plot, (c) Return level plot, (d) Density plot.}
\label{fig12}
\end{figure*}

The GEV provides a model for the distribution of block maxima. In the proposed method, the test metric $\Psi(\textit{\textbf{Z}}_{K},\hat{\textit{\textbf{X}}}_{\textrm{CRPFB}})$ is naturally the block maxima following (\ref{eq18}). The block size is the number of CRPFs used in the CRPFB. We will check the GEV model for the test metrics on $H_{0}$. Fig.~13 shows the diagnostic plots of the test metric of the second simulation on $H_{0}$, where the prior information of the possible signal is $f(t)\in[-80,80]~\textrm{Hz}$ and $f^{\prime}(t)\in[-120,120]~\textrm{Hz/s}$, and 2,000 CRPFs and 1 particle for each CRPF are used.
The diagnostic plots can be used to compare with the probability function, the quantile, the return level, and the probability density plotted on empirical and the GEV models. In Fig.~13, the empirical plots are obtained from $10^{5}$ test metrics in (\ref{eq16}) on $H_{0}$, and the parameters for the GEV model are estimated from only $1,500$ test metrics in (\ref{eq16}) on $H_{0}$.

Fig.~13(a) shows the empirical and GEV model-based distribution functions.
The empirical distribution function is estimated by Monte Carlo simulation from test metrics $\Psi_{m_{c}}(\textit{\textbf{Z}}_{K},\hat{\textit{\textbf{X}}}_{\textrm{CRPFB}})$ with $m_{c}=1,\ldots, M_{c}$ on $H_{0}$, and $M_{c}=10^{5}$. The GEV model-based distribution function is calculated with Equation~(\ref{eq16}), where the GEV parameters are estimated from 1,500 test metrics on $H_{0}$.
It can be seen from Fig.~13(a) that the empirical probabilities and GEV model-based probabilities are almost equal, and the line consisting of the two types of probabilities lies close to the unit diagonal. It means that the GEV model works well.

Fig.~13(b) shows the GEV model-based quantiles and the empirical quantiles,
where $\hat{G}^{-1}(m_{c}/(M_{c}+1))$ is the GEV model-based quantile corresponding to the probability $m_{c}/(M_{c}+1)$, and $\Psi_{m_{c}}(\textit{\textbf{Z}}_{K},\hat{\textit{\textbf{X}}}_{\textrm{CRPFB}})$ denotes the empirical quantile corresponding to the probability $m_{c}/(M_{c}+1)$.
It can be seen from Fig.~13(b) that the GEV model-based quantiles and the empirical quantiles are almost equal, and the line consisting of the two types of probabilities lies close to the unit diagonal. It also illustrates that the GEV model works well.

Fig.~13(c) shows the GEV model-based return level $\hat{G}^{-1}(m_{c}/(M_{c}+1))$ against $\textrm{log}(-\textrm{log}(1-m_{c}/(M_{c}+1)$, and empirical return level $\Psi_{m_{c}}(\textit{\textbf{Z}}_{K},\hat{\textit{\textbf{X}}}_{\textrm{CRPFB}})$ against $\textrm{log}(-\textrm{log}(1-m_{c}/(M_{c}+1)$. It can be seen from Fig.~13(c) that the GEV model-based curve and the empirical estimates are in reasonable agreement. It suggests that the GEV model is adequate.

Moreover, the density function plot in Fig.~13(d) is a comparison of an empirical probability density function based on the histogram of the test metrics on $H_{0}$ and the model-based probability density function based on the GEV model.

Fig.~13 shows that all the diagnostic plots provide support to the use of the GEV model. It means that the GEV model with the parameters estimated from fewer test metrics on $H_{0}$ can be used to describe the distribution of the test metrics on $H_{0}$. For example, in Fig.~13, the GEV model is estimated from only 1,500  test metrics on $H_{0}$, and the model fits well to the empirical distribution estimated from $10^5$ test metrics on $H_{0}$. By this way, the distribution of the test metrics on $H_{0}$ can be estimated from fewer observation data and the detection threshold can easily be expanded.

Based on the GEV model estimated from fewer test metrics on $H_{0}$, Fig.~14 compares the GEV model-based thresholds and the empirical thresholds. The empirical thresholds in Fig.~14 are estimated by Monte Carlo simulations using $10^5$ test metrics on $H_{0}$. The GEV model-based thresholds in Fig.~14 are the means of thresholds estimated from a different number of the test metrics on $H_{0}$. Each mean of the GEV model-based thresholds is calculated from 1,000 points.

\begin{figure*}[!htbp]
\setlength{\abovecaptionskip}{0pt}
\setlength{\belowcaptionskip}{0pt}
\centering
\includegraphics[width=6in,angle=0]{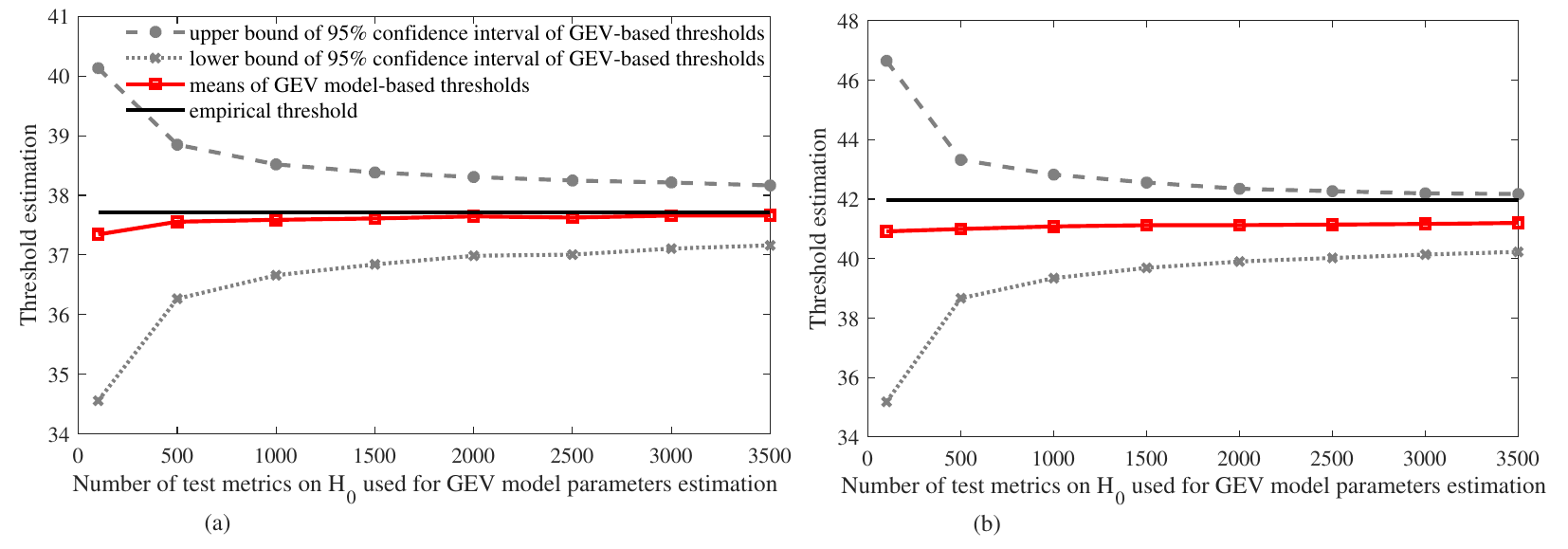}
\caption{Comparison of GEV model-based threshold and empirical threshold, (a) Comparison of GEV model-based threshold and empirical threshold for $P_{fa}=10^{-2}$, (b) Comparison of GEV model-based threshold and empirical threshold for $P_{fa}=10^{-3}$.}
\label{fig13}
\end{figure*}

It can be seen from Fig.~14(a) that, in the case of $P_{fa}=10^{-2}$, the empirical threshold and the GEV model-based threshold are very close to each other. And the model-based thresholds are stable when the parameters of GEV are estimated from more than 1,500 test metrics on $H_{0}$. The results in Fig.~13(a) show that the GEV model fits well with the test metrics on $H_{0}$.

However, Fig.~14(b) indicates that, in the case of $P_{fa}=10^{-3}$, there are some differences between the empirical threshold and the GEV model-based thresholds. The reason is that the number of the test metrics on $H_{0}$ used for empirical estimation is not large enough. In the case of $95\%$ confidence level of $P_{fa}=10^{-2}$, the empirical threshold estimated from $10^{5}$ test metrics on $H_{0}$ has about $6\%$ relative error. To achieve a similar confidence level and relative error, the empirical threshold with $P_{fa}=10^{-3}$ requires about $10^6$  test metrics on $H_{0}$. Therefore, the GEV model can help dramatically reduce the number of the test metrics on $H_{0}$ for threshold estimation.

\section{Conclusion}
This paper proposes an entirely parallel architecture of CRPF, i.e., CRPFB, for weak target detection. The problem of weak target detection is converted to a two-layer hypotheses test. The first layer of the hypotheses test is to estimate the IF curve of a possible target, which is implemented by the CRPFB. In this step, many types of detailed prior information are hypothesized. The CRPFs using different prior information are implemented in parallel and the output of the CRPF with the minimum cumulated cost is taken as the IF curve estimation results for the CRPFB. The second layer of the hypotheses test is to determine whether a target is present or not. In this step, the test metrics proportional to GLR and based on the IF curve estimation provided by the CRPFB is obtained and is compared with a given threshold. A target is declared to be present if the test metrics exceeds the threshold. Simulation results illustrate that the proposed CRPFB based detector has high detection capability, superior estimation capability, and low computational complexity. Moreover, the proposed method for threshold estimation fits well to the GEV model. Thus the GEV model estimated from fewer  $H_{0}$ can be used for threshold estimation. In the future, the proposed method will be combined with existing interest point detection methods~\cite{zhang2019corner2,zhang2019discrete} for improving the performance of different computer vision tasks such as image matching~\cite{zhang2020corner} and 3D reconstruction~\cite{zhang2017noise, zhang2023image, jing2022image,jing2022recent}

\ifCLASSOPTIONcaptionsoff
  \newpage
\fi



\bibliographystyle{IEEEtran}
\bibliography{myreference}
\begin{IEEEbiography}[{\includegraphics[width=1in,height=1.25in,clip,keepaspectratio]{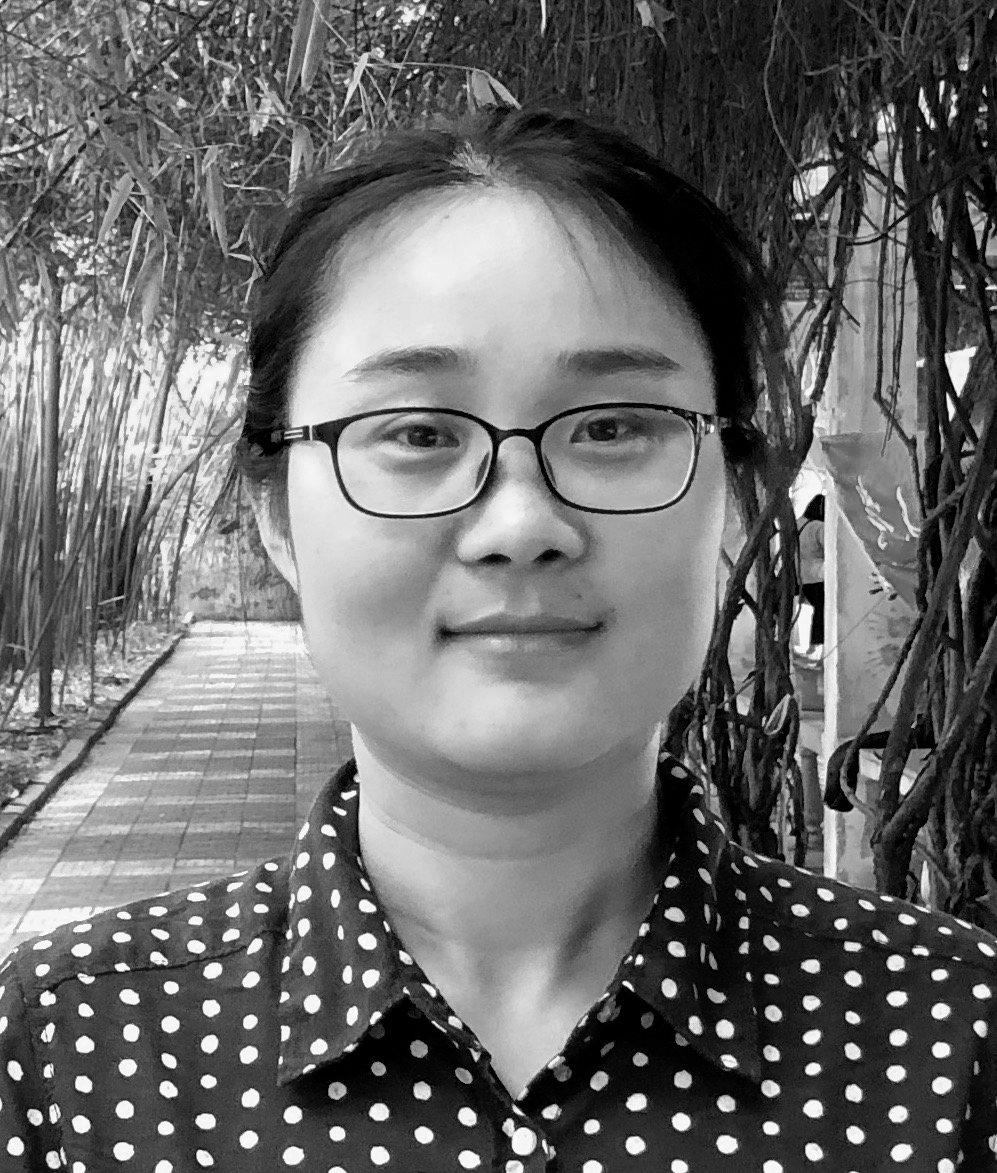}}]{Jin Lu}
was born in China in 1984. She received the B.E., M.E., and Ph.D degrees in signal and information processing from Xidian University, Xi'an, China, in 2007, 2010, and 2014 respectively.
She is currently a lecturer with the School of Electronic Information and Artificial Intelligence, Shaanxi University of Science \& Technology.
Her research interests include estimation,  tracking, signal processing, with an emphasis on weak target detection and tracking.
\end{IEEEbiography}
\vspace{-10 mm}

\begin{IEEEbiography}[{\includegraphics[width=1in,height=1.25in,clip,keepaspectratio]{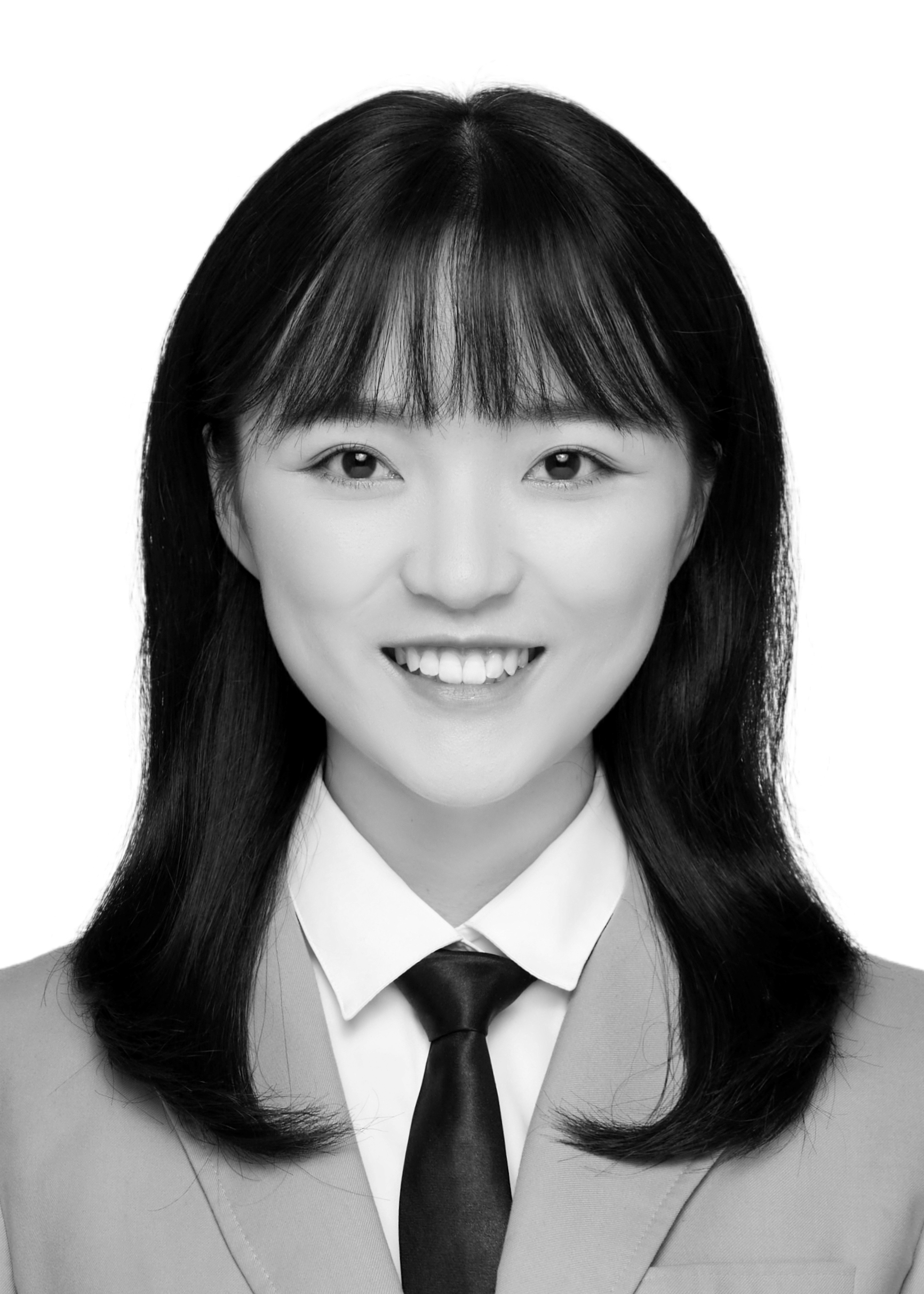}}]{Guojie Peng}
is a undergraduate student in new generation of computer science, Shaanxi University of Science \& Technology.
Her research interests are digital signal processing.
\end{IEEEbiography}
\vspace{-10 mm}

\begin{IEEEbiography}[{\includegraphics[width=1in,height=1.25in,clip,keepaspectratio]{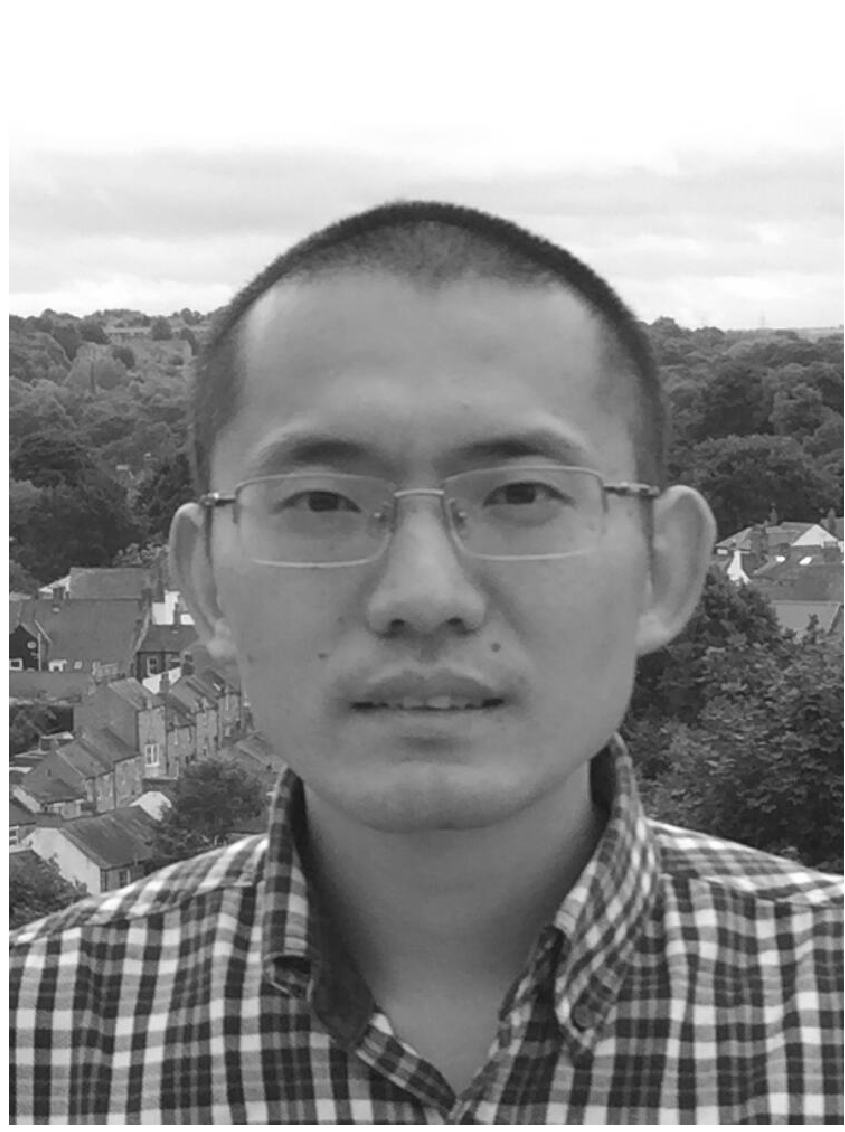}}]{Weichuan Zhang}
received the MS degree in signal and information processing from the Southwest Jiaotong University in China and the PhD degree in signal and information processing in National Lab of Radar Signal Processing, Xidian University, China. He is a research fellow at Griffith University, QLD, Australia. His research interests include computer vision, image analysis, and pattern recognition. He is a member of the IEEE.
\end{IEEEbiography}
\vspace{-10 mm}
\begin{IEEEbiography}[{\includegraphics[width=1in,height=1.25in,clip,keepaspectratio]{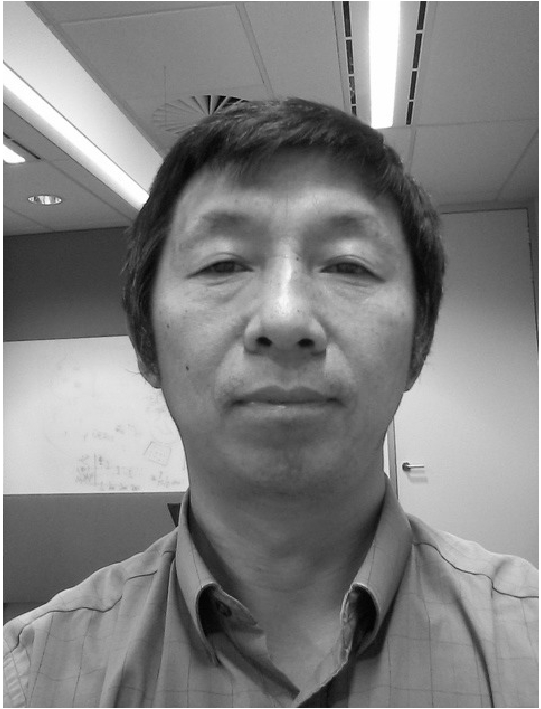}}]{Changming Sun}
received his PhD degree in computer vision from Imperial College London, London, UK in 1992. He then joined CSIRO, Sydney, Australia, where he is currently a Principal Research Scientist carrying out research and working on applied projects. He has served on the program/organizing committees of various international conferences. He is an Associate Editor of the EURASIP Journal on Image and Video Processing. His current research interests include computer vision, image analysis, and pattern recognition.
\end{IEEEbiography}
%\newpage

%\vfill



\end{document}